\documentclass[a4paper,11pt]{article}
\usepackage{jheppub}
\usepackage{amsmath,amssymb,bm,array,multirow,comment,slashed,epsfig}
\usepackage{feynmp}

\def \nc {N_c}

\newcommand{\Tint}[1]{{\hbox{$\sum$}\!\!\!\!\!\!\!\int\,}_{\!\!\!\!\raise-0.9ex\hbox{$\scriptstyle{#1}$}}}

\def \cc{{\mathcal C}}

\def\siml{{\ \lower-1.2pt\vbox{\hbox{\rlap{$<$}\lower6pt\vbox{\hbox{$\sim$}}}}\ }}
\def\simg{{\ \lower-1.2pt\vbox{\hbox{\rlap{$>$}\lower6pt\vbox{\hbox{$\sim$}}}}\ }}

\def \als {\alpha_{\mathrm{s}}}

\def \m2   {\mu^{2 \epsilon}}

\def\siml{{\ \lower-1.2pt\vbox{\hbox{\rlap{$<$}\lower6pt\vbox{\hbox{$\sim$}}}}\ }}
\def\simg{{\ \lower-1.2pt\vbox{\hbox{\rlap{$>$}\lower6pt\vbox{\hbox{$\sim$}}}}\ }}

\def\nn {\nonumber}

\def\xp {x_\perp}

\def\qp {q_\perp}

\def\bqp {\mathbf{q}_\perp}
\def\bxp {\mathbf{x}_\perp}

\def\2to2{{2\leftrightarrow 2}}

\def\LO{{\mathrm{LO}}}

\def\nfour{\mathcal{N}=4}
\def\qhat{\hat{q}}

\def\Fig#1{Fig.~\ref{#1}}

\def\q{{\bf{q}}}

\def\p{{\bf{p}}}
\def\x{{\bf{x}}}
\def\OO{{\mathcal{O}}}

\def\nfd{n_{F}}
\def\nbe{n_{B}}

\newcommand{\VEV}[1]{\left\langle{#1}\right\rangle}
\newcommand{\vev}[1]{\langle{#1}\rangle}

\newcommand{\minifeyn}[1]{\begin{minipage}{0.13\textwidth}
\begin{fmfgraph}(48,60)\fmfpen{0.2thin}\fmfset{dot_len}{0.8mm}
#1
\end{fmfgraph}\end{minipage}}

\newcommand{\minifeynLO}[1]{\begin{minipage}{0.18\textwidth}
\begin{fmfgraph}(70,50)\fmfpen{0.2thin}\fmfset{dot_len}{0.8mm}
#1
\end{fmfgraph}\end{minipage}}

\title{Transverse momentum broadening and collinear radiation at NLO in the $\mathcal{N}$=4 SYM plasma}
\author[1]{Jacopo Ghiglieri,}
\author[1,2]{HyungJoo Kim}

\affiliation[1]{Theoretical Physics Department, CERN,\\ CH-1211 Geneva 23, Switzerland}
\affiliation[2]{Department of Physics, Yonsei University,\\ 50 Yonsei-ro, Seoul 03722, Korea} 

\emailAdd{jacopo.ghiglieri@cern.ch}
\emailAdd{hyungjoo.kim@cern.ch}
\preprint{CERN-TH-2018-181}
\abstract{
We compute $\OO(g)$ NLO corrections to the transverse scattering kernel and transverse momentum
broadening coefficient $\qhat$ of weakly-coupled $\nfour$ SYM. Based on this, we also compute
NLO correction to the collinear splitting rates. For $\qhat$ we find that the NLO/LO ratio is
similar to the QCD one, with large NLO corrections. This is contrasted by our findings
for the collinear splitting rate, which show a much better convergence in SYM
than in QCD, providing further support to earlier expectations that NLO 
corrections have signs and relative magnitudes controlled by the specifics of the theory.
We also compare the ratio of $\qhat$
in QCD and in $\nfour$ theory to strong coupling expectations. 

}


\begin{document}

\maketitle
\flushbottom

\section{Introduction}
The characterization of the QCD medium produced in heavy ion collisions, and possibly
also in smaller systems (proton-nucleus and high-multiplicity proton-proton collisions)
proceeds through the complementary study of  bulk properties and  hard probes.
The AdS/CFT correspondence \cite{Maldacena:1997re,Witten:1998qj,Gubser:1998bc}
has been widely applied in both cases over the past 20 years (see
 \cite{CasalderreySolana:2011us} for a review), providing great
 qualitative insight on the strong-coupling regime of QCD,
in particular for quantities such as the specific shear viscosity $\eta/s$ 
\cite{Policastro:2001yc,Kovtun:2003wp,Kovtun:2004de}, which are not directly
accessible from lattice QCD without arduous analytical continuations (see
\cite{Meyer:2011gj} for a review).

In its standard form, the AdS/CFT correspondence conjectures a duality
between conformal $\nfour$ Super-Yang-Mills (SYM) theory in $D=4$ spacetime dimensions
and  type-IIB string theory in $AdS_5\times S^5$. At large
values of the 't Hooft coupling $\lambda$ and for large numbers of colors $N_c$,
calculations on the holographic side become accessible 5D gravity computations; 
a finite-temperature system in the CFT corresponds to a black hole in AdS. This has lead
to the aforementioned wealth of computations of quantities of interest for heavy-ion physics at strong
coupling. If one wants more than a qualitative insight when applying 
strong-coupling, holographic results to the QCD medium, one needs to understand
a double extrapolation: from $\nfour$ SYM to QCD, and from $\lambda\to\infty$
to the regime of ``intermediate'' couplings one expects for heavy ion collisions,
such as $\als\equiv g^2/(4\pi)\approx0.3$, $\lambda\equiv g^2N_c\approx10$.

To this end, an important step forward would be the understanding of thermal
$\nfour$ SYM at all values of $\lambda$. On the large coupling side,
for many quantities,
such as the photon rate \cite{CaronHuot:2006te} or the viscosity,
the first corrections in the inverse coupling in AdS have been computed,
in \cite{Hassanain:2012uj} and  \cite{Buchel:2004di,Benincasa:2005qc,Buchel:2008ac,Buchel:2008wy}
respectively. 
At weak coupling, calculations can be performed directly on the CFT side, using the tools
of finite-temperature perturbation theory. Leading-order (LO) results for the photon
rate and the viscosity have been presented in \cite{CaronHuot:2006te} and
\cite{Huot:2006ys}, in both cases following the path set by previous
perturbative QCD (pQCD) calculations, \cite{Arnold:2001ms} and \cite{Arnold:2000dr,Arnold:2003zc}
respectively. Once  $\nfour$ SYM results in the two regimes are available,
one can try to extrapolate from both sides towards the interesting intermediate
region.  

Clearly, this exercise requires the best possible knowledge on both sides. At weak
coupling, calculations of transport coefficients and of dynamical quantities
like the photon rate are notoriously difficult, requiring different sets 
of resummations already at LO. Over the past decade, a new understanding
of the analytical properties of thermal amplitudes at light-like separations
has emerged \cite{CaronHuot:2008ni} (see \cite{Ghiglieri:2015zma} for a more
pedagogic presentation). Owing to this development,
next-to-leading order (NLO) pQCD calculations of the photon rate \cite{Ghiglieri:2013gia}
and of transport coefficients \cite{Ghiglieri:2018dib,Ghiglieri:2018dgf} have recently
appeared. Their extension to $\nfour$ SYM would thus lead 
to an  ``extrapolation game'', as we have sketched before, with the first corrections in
both regimes and thus a first estimate of the uncertainty. Furthermore, one could compare
the QCD and SYM NLO weak-coupling results, to get a better understanding on how to deal with the different
type and number of degrees of freedom when applying holographic results.

Motivated by this, in this paper we investigate the \emph{jet quenching parameter} $\qhat$
at NLO in $\nfour$ SYM. Also known as the transverse
momentum broadening coefficient, $\qhat\equiv <p_\perp^2>/L$ describes how much transverse momentum
$p_\perp$ is picked up per length $L$ by a highly energetic parton propagating through a plasma.
It is thus extremely important for the physics of jet quenching (see 
\cite{d'Enterria:2009am,Wiedemann:2009sh,Majumder:2010qh,Mehtar-Tani:2013pia,Roland:2014jsa,Qin:2015srf,Connors:2017ptx}),
where it is a parameter in many theoretical models of medium-modified radiation from
the hard partons constituting the jet (see \cite{Burke:2013yra} for a comparison with/extraction
from data and \cite{Armesto:2011ht} for a theoretical comparison of the models). 
In this paper, we will study $\qhat$ and the related \emph{transverse scattering kernel} (or
collision kernel)
$\cc(\qp)$ at NLO. Both quantities are known in the strong-coupling limit \cite{Liu:2006ug,Liu:2006he,D'Eramo:2010ak},
together with part of the first corrections in the inverse coupling \cite{Armesto:2006zv}
(see also \cite{Gubser:2006nz,CasalderreySolana:2007qw} for
related strong-coupling calculations for massive probes). In weak-coupling
QCD, leading- and next-to-leading results have been obtained in \cite{Aurenche:2002pd,Arnold:2008vd} and
\cite{CaronHuot:2008ni} respectively. Hence, we shall be in the position to compare
both weak-coupling results in the two different theories and weak and strong coupling within $\nfour$ SYM.

Furthermore, $\qhat$ and $\cc(\qp)$ are two of the main ingredients in many other perturbative calculations:
the former has been found in \cite{Ghiglieri:2018dib} to be the main driver of the large NLO corrections
to the transport coefficients, while the latter determines the NLO correction to \emph{collinear radiation},
which is an important ingredient both in the kinetic theory used to determine transport coefficients
\cite{Arnold:2002zm,Ghiglieri:2015ala} and in the thermal photon rate. Hence, as a first application
of our NLO results, we also compute the collinear radiation rate, or 
collinear splitting rate, at NLO, to study its sensitivity 
to the different quasiparticle degrees of freedom in QCD and SYM.

At the technical level, the calculation
of $\qhat$ and $\cc(\qp)$ can be separated into
soft ($gT$-scale) and hard ($T$-scale) contributions. At leading-order the
soft contribution is known \cite{Aurenche:2002pd,CaronHuot:2006te},
while in the hard region we will determine the 
contribution from the SYM scalars.
At NLO only the soft scale enters:
our  calculation 
 requires the evaluation of soft one-loop corrections to the
Wilson loop from which these are defined \cite{D'Eramo:2010ak,Benzke:2012sz}. The
aforementioned new understanding of the analytical properties of thermal amplitudes  leads  in
this case to a great simplification, as the problem can be mapped to a much simpler one within 
the dimensionally-reduced Euclidean theory. In the case of QCD this is Electrostatic QCD (EQCD)
\cite{Braaten:1994na,Braaten:1995cm,Braaten:1995jr,Kajantie:1995dw,Kajantie:1997tt},
while in the case of SYM this is usually called ESYM \cite{Nieto:1999kc}, whose kinetic
term was written down in \cite{VazquezMozo:1999ic,Kim:1999sg,Nieto:1999kc}. We will thus need to analyze in
detail the contribution of the scalars of $\nfour$ theory to ESYM in general and to our observables
in particular (fermions are not an explicit degree of freedom of the dimensionally-reduced
theories). 

Finally, we also remark that the transverse scattering kernel of the electroweak (EW)
theory is similarly
a very important ingredient in determinations of the collinear radiation rate within that theory,
which is of relevance for applications such as the collinear thermal production of right-handed 
neutrinos \cite{Anisimov:2010gy,Besak:2012qm}. Our NLO calculation, with its
in-depth analysis of the scalar contribution, will thus be very helpful in extending the EW
calculation towards NLO, whose necessity has been pointed out in \cite{Ghiglieri:2016xye} and 
which requires the evaluation of the contribution of the Higgs scalar doublet.

The paper is organized as follows: in Sec.~\ref{ESYM} we review the theoretical setup
and definitions. Sec.~\ref{sec_collisionkernel} is devoted to the computation
of $\cc(\qp)$ at NLO in ESYM, while in Sec.~\ref{sec_qhat} we discuss
$\qhat$ at NLO and its relation with the strong-coupling and QCD results.
In Sec.~\ref{sec_collinear} we apply our results to the collinear splitting
rate and we draw our conclusions in Sec.~\ref{sec_concl}. Technical details
are to be found in the Appendices.

\section{Theoretical setup}\label{ESYM}

\begin{figure}[ht]
 \centering
 \includegraphics[width=0.55\textwidth]{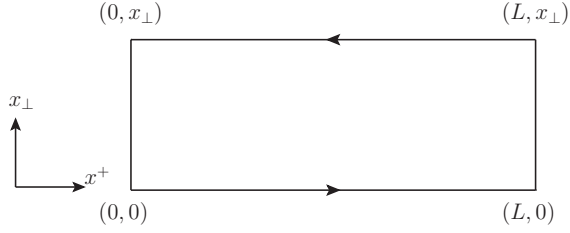}
 \caption{Wilson loop $W(\xp,L)$ leading to $\mathcal{C}'(x_\perp)$.}
 \label{wilsonloop}
 \end{figure}

The collision kernel $\mathcal{C}(q_\perp)$ describing the evolution of the transverse momentum of a very hard particle
with momentum $\p$ and energy $E\approx p$, with $E\gg T$, is
defined as 
\begin{align}
\mathcal{C}(q_\perp)\equiv \lim_{p\to\infty}(2\pi)^2\frac{d\Gamma_{\text{scatt}}(\p,\p+\bqp)}{d^2q_\perp},
\end{align}
where $q_{\perp}$ is the transverse momentum ($\q_\perp\cdot\p=0$) acquired in the scattering.\footnote{In our convention
the metric is $g_{\mu\nu}=(-+++)$, $P=(p^0,\p)$ is a four-vector, with $\p$  the three-vector with modulus $p$.}
The collision kernel can be defined in a field-theoretical manner
using the Wilson loop in the $(x^+,x_\perp)$ plane \cite{D'Eramo:2010ak,Benzke:2012sz} sketched in
 Fig.~\ref{wilsonloop}.
  Here we use light-cone coordinates defined as $p^-\equiv p^0-p^z$, $p^+\equiv\frac{p^0+p^z}{2}$, and $\p_\perp=(p^x,p^y)$ for convenience. 
The Wilson loop is then written explicitly as
\begin{equation}
	\VEV{W(\xp,L)}=\frac{1}{d_R}\bigg\langle\mathrm{Tr}\,\tilde U(0,\x_\perp;0)U(0,L;\x_\perp)
	\tilde U(\x_\perp,0;L) U(L,0;0)\bigg\rangle,
	\label{defwloop}
\end{equation}
where the Wilson lines are defined as\footnote{This definition is valid
both in QCD and in $\nfour$ SYM.}
\begin{align}
	U(a,b;x_\perp)&=P\exp\left(-ig\int_b^a dx^+ A^-(x^+,x^-=0,x_\perp)\right),\nn\\
	\tilde U({\bf a},{\bf b};x^+)&=P\exp\left(-ig\int_0^1 ds\, ({\bf a}-{\bf b})\cdot {\bf A}_\perp(x^+,x^-=0,x_\perp=
	s({\bf a}-{\bf b}))\right).
\end{align}
The trace runs over the color degrees of freedom, with the source taken in representation $R$
with $d_R$ its dimension. In QCD, hard quarks will be described by  the fundamental Wilson 
loop $R=F$ and hard gluons by the adjoint one, $R=A$. In $\nfour$ SYM all d.o.f.s transform
in the adjoint. $\langle\ldots\rangle$ denotes a thermal average. Finally, the
operators should be intended ordered such that fields on the backward-propagating
Wilson line $U(0,L;\x_\perp)$ come always to the left of those in the forward-propagating one, as they
are associated with the conjugate amplitude and amplitude respectively or, in the Schwinger-Keldysh
language, the anti-time ordered and time-ordered parts of the contour \cite{D'Eramo:2010ak,Benzke:2012sz}.

$\cc(\qp)$ can then be obtained from this Wilson loop in the large $L$ limit as
\begin{align}
\mathcal{C}'(\xp)=-\lim_{L\to\infty}\frac{1}{L}\ln\,\VEV{W(\xp,L)}
\end{align}
where 
$\mathcal{C}'(\xp)$ is the collision kernel in impact-parameter space, i.e.
\begin{align}
	\label{defcxp}
\mathcal{C}'(\xp)\equiv\int \frac{d^2\qp}{(2\pi)^2}\,(1-e^{i\bqp\cdot\bxp})\,\mathcal{C}(\qp).
\end{align}
We use the $\mathcal{C}'(\xp)$ notation as a reminder that $\mathcal{C}'(\xp)$ \emph{is not} the Fourier transform
of $\mathcal{C}(\qp)$.\footnote{\label{foot_prob}
Equally importantly, we remark that $\mathcal{C}(\qp)$ \emph{is not} 
the Fourier transform of $\VEV{W(\xp,L)}$. That quantity is instead called $P(\qp)$, 
 the probability for the hard parton to pick up a certain transverse momentum $\qp$. It reads
 \cite{D'Eramo:2010ak,Benzke:2012sz}
\begin{equation}
	\label{defp}
	P(\qp)\equiv\int d^2\xp e^{-i\bqp\cdot\bxp}\VEV{W(\xp,L)}.
\end{equation}
}

Finally, the transverse momentum broadening coefficient $\qhat$ can be obtained as the second
moment of $\cc(\qp)$, i.e.
\begin{equation}
	\label{defqhat}
	\qhat\equiv \int^{q_\mathrm{\max}}_0 \frac{d^2\qp}{(2\pi)^2}\,\qp^2\,\cc(\qp)\,,
\end{equation}
where $q_\mathrm{max}$ is a process-dependent UV regulator that is in general needed at weak coupling. As we
are interested in hard particles of energy/momentum $E\gg T$, we shall take $q_\mathrm{max}\gg T$
as well.\footnote{Our definition, Eq.~\eqref{defqhat},
 is slightly different from the one in \cite{D'Eramo:2010ak,Benzke:2012sz},
which involves the second moment of $P(\qp)$, defined in Footnote~\ref{foot_prob}. However,
as shown in \cite{DEramo:2012uzl}, the two different definitions give rise to the
same result.} Finally, the eikonal propagation of the hard parton, which gives rise to the
Wilson loop above, is only valid insofar the transferred momentum $\qp$ does not affect the hard parton's. 
Hence, we must also require $E\gg q_\mathrm{max}$.

At leading order the scattering kernel and $\qhat$ receive contributions from
gluon-mediated elastic scatterings off the medium constituents \cite{CaronHuot:2008ni,Arnold:2008vd}. 
This is true in any gauge theory. Since the Coulomb scattering matrix element squared
is proportional to $t^{-2}$, we expect a $1/\qp^4$ behavior in the scattering kernel. Thus
scatterings with $\qp\sim T$ and softer ones with $\qp\sim gT$ contribute at the same order
(up to logarithms) to Eq.~\eqref{defqhat}. $gT$ is the scale at which collective effects
such as Debye screening and Landau damping appear in a weakly-coupled plasma; we call it the 
\emph{soft scale}. We will call $T$ the \emph{hard} scale, with the understanding that the energy/momentum
$E\gg T$ of the probe is called \emph{very hard}. It is convenient to introduce a regulator
$q^*$, with $gT\ll q^*\ll T$, to separate the soft and hard contributions, i.e.
\begin{equation}
\hat{q}= \int^{q^*} _0 \frac{d^2q_\perp}{(2\pi)^2}\,q_\perp^2\, \mathcal{C}_\mathrm{soft}(q_\perp)+\int^{q_\mathrm{max}}_{q^*} \frac{d^2q_\perp}{(2\pi)^2}\,q_\perp^2\, \mathcal{C}_{\rm{hard}}(q_\perp),
\label{splitqhat}
\end{equation}

In perturbative Thermal Field Theory, soft bosons in equilibrium are highly occupied, since the
Bose--Einstein distribution $\nbe(q^0)=(e^{q^0/T}-1)^{-1}$ becomes approximately $T/q^0\sim 1/g$ there.
Hence, soft-boson loops are not suppressed by the usual factor of $g^2$, which characterizes hard (or zero-temperature)
loops, but by a single factor of $g$. Therefore, NLO corrections to $\qhat$
come from one-loop bosonic diagrams at $\qp\sim gT$. In QCD these are only gauge (gluon and ghost) loops,
whereas in $\nfour$ SYM there is also a scalar contribution. In either case,
the analytical properties stemming from light-cone causality allow these contributions
to $\cc_\mathrm{soft}(\qp)$ (and thereupon $\qhat$) to be computed within the simpler dimensionally-reduced theory, as we
show in the next section. We will then discuss $\qhat$ and 
$\cc_\mathrm{hard}(\qp)$ in Sec.~\ref{sec_qhat}. Finally, we remark that, by the same arguments
on the Bose--Einstein distribution, the \emph{ultrasoft} (or magnetic) 
scale $g^2T$ has no loop suppression factor and is thus
non-perturbative \cite{Linde:1980ts}. 
However, its contribution to $\qhat$ is suppressed by $g^2$ with respect to the LO
 \cite{CaronHuot:2008ni,Benzke:2012sz,Laine:2012ht,DEramo:2012uzl} and 
is thus beyond our $\OO(g)$ accuracy. Knowledge of $\mathcal{C}(\qp\sim g^2T)$ is also not needed, as it does
not contribute to the collinear radiation rates at LO or NLO. As remarked in 
 \cite{Benzke:2012sz,Laine:2012ht,Panero:2013pla},
both can be obtained within the Euclidean framework using non-perturbative methods.

\section{The collision kernel at NLO}
\label{sec_collisionkernel}
In principle, a one-loop, soft-loop calculation would require a thorough analytical and numerical effort
in the Hard Thermal Loop (HTL) effective theory \cite{Braaten:1989mz,Frenkel:1989br}
(see \cite{Czajka:2012gq} for the complete HTL structure of $\nfour$ SYM).
However, as shown by Caron-Huot \cite{CaronHuot:2008ni} in the context of the NLO calculation
 of $\qhat$ in QCD,
light-cone causality makes  correlators such as Eq.~\eqref{defwloop} dramatically simpler, 
leading to a Euclidean formulation. 
Indeed, he  has shown that the soft contribution to many space-like separated correlators
can be mapped to a dimensionally reduced Euclidean theory, which integrates out all the non-zero Matsubara modes.
Based on that, he proceeded to determine the one-loop, soft-loop contribution to Eq.~\eqref{defwloop}
with EQCD.

We can apply the same strategy to $\nfour$ SYM theory.
Electrostatic SYM \cite{VazquezMozo:1999ic,Kim:1999sg,Nieto:1999kc} 
is a dimensionally-reduced 3D Euclidean theory of $\nfour$ SYM for the soft $(gT)$ scale physics. 
We  now use this effective theory to calculate the $\OO(g)$ contribution to 
the collision kernel and  to the transverse momentum broadening coefficient 
of the weakly coupled $\nfour$ SYM plasma. 
Keeping only the operators that are needed for this calculation, the Lagrangian of
ESYM reads
\begin{align}
\mathcal{L}_{ESYM}&=\frac{1}{4}F^a_{ij}F^a_{ij}+\frac{1}{2}(D_iA_0)^a(D_iA_0)^a+\frac{1}{2}(D_i\phi_I)^a(D_i\phi_I)^a
+\frac{1}{2}m_E^2A_0^a A_0^a+\frac{1}{2}m_S^2\phi_I^a \phi_I^a\nonumber\\
&+\frac{1}{2}\lambda_{ES}f^{abc}f^{ade}\phi_I^bA_0^c\phi_I^dA_0^e+\ldots,
\label{esymlag}
\end{align}
where $F^a_{ij}=\partial_iA^a_j-\partial_jA^a_i+g_Ef^{abc}A^b_iA^c_j$ and 
$(D_i\Phi)^a=\partial_i\Phi^a+g_Ef^{abc}A_i^b\Phi^c$ for any SU($N_c$) 
adjoint bosonic field $\Phi^a\in\{A_i^a,A_0^a,\phi_I^a \}$. $I\in{1,\ldots,n_S}$ spans the $n_S=6$ 
real scalars of $\nfour$ SYM. The interaction term on the second line is new to this work, as the references
in the literature focused on the kinetic terms only. These reflect how,
under dimensional reduction, the spatial components $A_i^a$ remain gauge fields, 
while $A_0^a$ becomes a massive, \emph{electrostatic} 
adjoint scalar field and the scalars $\phi^a_I$ become massive as well,
with masses $m_E$ and $m_S$ respectively. Fermions are fully integrated out during dimensional reduction
and only contribute to the Wilson coefficients, such as these masses.
At leading order, the matching between $\nfour$ SYM and ESYM yields  \cite{VazquezMozo:1999ic,Kim:1999sg,Nieto:1999kc}
\begin{align}
g_E^2&=g^2T,\qquad \lambda_{ES}=g^2T,\label{matchcoupling}\\
\label{matchmasses}
m_E^2&=\frac{C_Ag^2T^2}{3}\left(1+\frac{n_S}{2}+\frac{n_f}{2}\right)=2\lambda T^2,\quad
 m_S^2=\frac{C_Ag^2T^2}{6}\left(1+\frac{n_S}{2}+\frac{n_f}{2}\right)=\lambda T^2,
\end{align}
where $g_E$ is the dimensionful coupling of ESYM, $g$ is the coupling of $\nfour$ SYM, $C_A=N_c$ the quadratic Casimir of the
adjoint representation and $\lambda\equiv g^2N_c$ is the 't Hooft coupling. 
In our conventions $n_f=4$ counts the number of Weyl fermion species of $\nfour$ SYM.
We note that the electrostatic mass $m_E$ equals at leading order the Debye mass; we
shall use both terms in the following.
Our graphical conventions for scalars and gauge fields, as well as the Feynman rules derived from
Eq.~\eqref{esymlag}, are illustrated in App.~\ref{app_feyn}.

\begin{figure}[ht]
\centering
\begin{fmffile}{diagram_LO}
\fmfset{curly_len}{1.5mm}
\fmfset{dot_len}{1.3mm}
\minifeynLO{\fmfstraight
\fmfleft{i1,i2}
\fmfright{o1,o2}
\fmf{vanilla}{i1,v1,o1}
\fmf{vanilla}{i2,v2,o2}
\fmffreeze
\fmf{vanilla}{v1,v3}
\fmf{vanilla}{v3,v2}\\ \vspace{0.5em}\hspace{2.2em}  (a)}
\minifeynLO{\fmfstraight
\fmfleft{i1,i2}
\fmfright{o1,o2}
\fmf{vanilla}{i1,v1,o1}
\fmf{vanilla}{i2,v2,o2}
\fmffreeze
\fmf{gluon}{v1,v2}
\\ \vspace{0.5em} \hspace{2.2em}  (b)}
\caption{Diagrams contributing to $\mathcal{C}_\mathrm{soft}(q_{\perp})$ at LO.  As shown in detail
in App.~\ref{app_feyn}, curly lines
are the $A^i$ gauge fields, solid lines are the electrostatic scalars $A^0$. 
The parallel lines represent the Wilson line along the $x^+$ direction separated by $x_\perp$.}
\label{fig_LO}
\end{fmffile}
\end{figure}
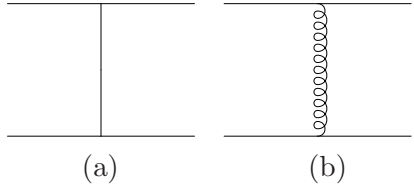
As we explained, the calculation of this Wilson loop is reduced to the calculation using ESYM.
In any non-singular gauge 
the transverse Wilson lines become irrelevant in the large-$L$ limit. There, the LO contribution
from the soft region comes from the single-gluon exchange diagrams shown in Fig.~\ref{fig_LO}.
We exploit the manifest gauge invariance
of the definition in Eq.~\eqref{defwloop} to compute in Feynman gauge, as shown in App.~\ref{app_feyn}
and \ref{app_detail}.  
A straightforward evaluation leads then to
\cite{CaronHuot:2006te,Aurenche:2002pd}
\begin{equation}
	\label{clo}
	\mathcal{C}^{(\text{LO})}_\mathrm{soft}(q_\perp)=g^2C_RT\left[\frac{1}{q_\perp^2}-\frac{1}{q_\perp^2+m_E^2}\right]
	 =\frac{g^2C_R T m_E^2}{q_\perp^2(q_\perp^2+m_E^2)},
\end{equation}
where the first term in square brackets is the gauge propagator $G^{zz}$ and the second term the massive,
electrostatic propagator $G^{00}$. Even though in $\nfour$ SYM all sources are adjoint, we prefer
for now to keep $C_R$ unassigned, so that we can keep the connection to the QCD calculation more transparent
and at the same time keep our results more general, so that they can be more easily adapted to the electroweak theory
as well, with its different group theory factors.

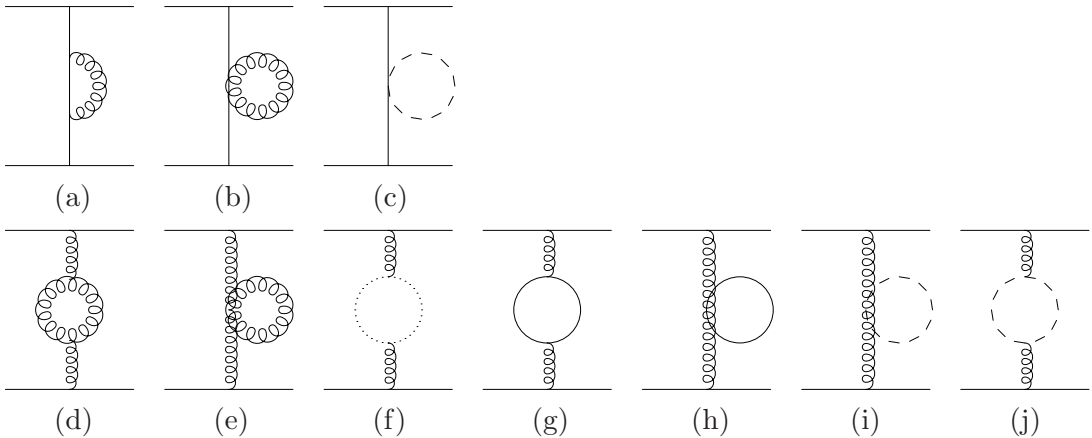
\begin{figure}[ht]
\begin{fmffile}{diagram_NLO}
\fmfset{curly_len}{1.5mm}
\fmfset{dot_len}{1.3mm}
\minifeyn{\fmfstraight
\fmfleft{i1,i2}
\fmfright{o1,o2}
\fmf{vanilla}{i1,v1,o1}
\fmf{vanilla}{i2,v2,o2}
\fmffreeze
\fmf{vanilla}{v1,v3}
\fmf{vanilla}{v3,v4}
\fmf{gluon,right,tension=0.005,tag=2}{v3,v4}
\fmf{vanilla}{v4,v2}\\ \vspace{0.5em} $\quad$ (a)}
\minifeyn{\fmfstraight
\fmfleft{i1,i2}
\fmfright{o1,o2}
\fmf{vanilla}{i1,v1,o1}
\fmf{vanilla}{i2,v2,o2}
\fmffreeze
\fmf{vanilla}{v1,v3}
\fmfi{gluon}{fullcircle scaled .35w shifted (.74w,.5h)}
\fmf{vanilla}{v3,v2}\\ \vspace{0.5em} $\quad$ (b)}
\minifeyn{\fmfstraight
\fmfleft{i1,i2}
\fmfright{o1,o2}
\fmf{vanilla}{i1,v1,o1}
\fmf{vanilla}{i2,v2,o2}
\fmffreeze
\fmf{vanilla}{v1,v2}
\fmfi{dashes}{fullcircle scaled .52w shifted (.76w,.5h)}\\ \vspace{0.5em} $\quad$ (c)}\\
\minifeyn{\fmfstraight
\fmfi{gluon}{fullcircle scaled .35w shifted (.5w,.5h)}
\fmfleft{i1,i2}
\fmfright{o1,o2}
\fmf{vanilla}{i1,v1,o1}
\fmf{vanilla}{i2,v2,o2}
\fmffreeze
\fmf{gluon}{v1,v3}
\fmf{phantom,left,tension=0.4,tag=1}{v3,v4}
\fmf{phantom,left,tension=0.4,tag=2}{v4,v3}
\fmf{gluon}{v4,v2}\\ \vspace{0.5em} $\quad$ (d)}
\minifeyn{\fmfstraight
\fmfleft{i1,i2}
\fmfright{o1,o2}
\fmf{vanilla}{i1,v1,o1}
\fmf{vanilla}{i2,v2,o2}
\fmffreeze
\fmf{gluon}{v1,v3}
\fmfi{gluon}{fullcircle scaled .35w shifted (.74w,.5h)}
\fmf{gluon}{v3,v2}\\ \vspace{0.5em} $\quad$ (e)}
\minifeyn{\fmfstraight
\fmfleft{i1,i2}
\fmfright{o1,o2}
\fmf{vanilla}{i1,v1,o1}
\fmf{vanilla}{i2,v2,o2}
\fmffreeze
\fmf{gluon}{v1,v3}
\fmf{dots,left,tension=0.35,tag=1}{v3,v4}
\fmf{dots,left,tension=0.35,tag=2}{v4,v3}
\fmf{gluon}{v4,v2}\\ \vspace{0.5em} $\quad$ (f)}
\minifeyn{\fmfstraight
\fmfleft{i1,i2}
\fmfright{o1,o2}
\fmf{vanilla}{i1,v1,o1}
\fmf{vanilla}{i2,v2,o2}
\fmffreeze
\fmf{gluon}{v1,v3}
\fmf{vanilla,left,tension=0.35,tag=1}{v3,v4}
\fmf{vanilla,left,tension=0.35,tag=2}{v4,v3}
\fmf{gluon}{v4,v2}\\ \vspace{0.5em} $\quad$ (g)}
\minifeyn{\fmfstraight
\fmfleft{i1,i2}
\fmfright{o1,o2}
\fmf{vanilla}{i1,v1,o1}
\fmf{vanilla}{i2,v2,o2}
\fmffreeze
\fmf{gluon}{v1,v2}
\fmfi{vanilla}{fullcircle scaled .52w shifted (.76w,.5h)}\\ \vspace{0.5em} $\quad$ (h)}
\minifeyn{\fmfstraight
\fmfleft{i1,i2}
\fmfright{o1,o2}
\fmf{vanilla}{i1,v1,o1}
\fmf{vanilla}{i2,v2,o2}
\fmffreeze
\fmf{gluon}{v1,v2}
\fmfi{dashes}{fullcircle scaled .52w shifted (.76w,.5h)}\\ \vspace{0.5em} $\quad$ (i)}
\minifeyn{\fmfstraight
\fmfleft{i1,i2}
\fmfright{o1,o2}
\fmf{vanilla}{i1,v1,o1}
\fmf{vanilla}{i2,v2,o2}
\fmffreeze
\fmf{gluon}{v1,v3}
\fmf{dashes,left,tension=0.35,tag=1}{v3,v4}
\fmf{dashes,left,tension=0.35,tag=2}{v4,v3}
\fmf{gluon}{v4,v2}\\ \vspace{0.5em} $\quad$ (j)}
\caption{One-loop self-energy diagrams contributing to  $\delta\mathcal{C}(q_{\perp})$. In addition
to the conventions of Fig.~\ref{fig_LO}, dotted lines are ghosts and dashed lines
are the scalars $\phi_I$.}
\label{fig_oneloop}
\end{fmffile}
\end{figure}
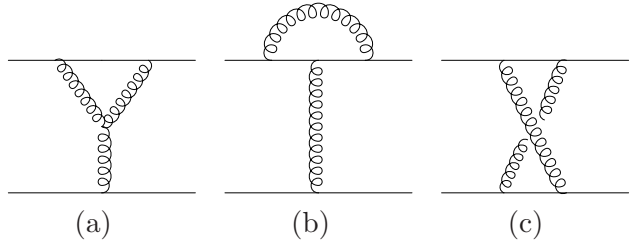
\begin{figure}[ht]
	\vspace{1cm}
\centering
\begin{fmffile}{diagram_NLO2}
\fmfset{curly_len}{1.5mm}
\fmfset{dot_len}{1.3mm}
\minifeynLO{\fmfstraight
\fmfleft{i2,i1}
\fmfright{o2,o1}
\fmf{vanilla}{i1,v1,v11,v111,o1}
\fmf{vanilla}{i2,v2,o2}
\fmffreeze
\fmf{gluon}{v3,v1}
\fmf{gluon}{v3,v111}
\fmf{gluon}{v2,v3}
\fmfforce{0.5w,0.5h}{v3}
\\ \vspace{0.5em} \hspace{1.6em}  (a)}
\minifeynLO{\fmfstraight
\fmfleft{i2,i1}
\fmfright{o2,o1}
\fmf{vanilla}{i1,v1,v11,v111,o1}
\fmf{vanilla}{i2,v2,o2}
\fmffreeze
\fmf{gluon,right,tension=0.001}{v111,v1}
\fmf{gluon}{v11,v2}
\\ \vspace{0.5em} \hspace{1.6em}  (b)}
\minifeynLO{\fmfstraight
\fmfleft{i2,i1}
\fmfright{o2,o1}
\fmf{vanilla}{i1,v1,v11,o1}
\fmf{vanilla}{i2,v2,v22,o2}
\fmffreeze
\fmf{gluon}{v1,v22}
\fmf{gluon}{v11,v3}
\fmf{phantom}{v3,v4}
\fmf{gluon}{v4,v2}
\fmfforce{0.55w,0.55h}{v3}
\fmfforce{0.46w,0.4h}{v4}
\\ \vspace{0.5em} \hspace{1.6em}  (c)}
\caption{Some ``multi-gluon-exchange'' diagrams contributing to NLO.}
\label{fig_mul}
\end{fmffile}
\end{figure}

The NLO soft scattering kernel can be written symbolically as 
\begin{equation}
	\label{defdc}
	\mathcal{C}^{(\text{NLO})}_\mathrm{soft}(q_\perp)\equiv\mathcal{C}^{(\text{LO})}_\mathrm{soft}(q_\perp)
	+\delta \cc(\qp)\,.
\end{equation}
The NLO contributions in $\delta \cc(\qp)$ can come from the soft one-loop self-energy insertion into  
the LO diagrams and from ``multi-gluon exchanges'' (e.g. two single-gluon exchanges, three gluon vertex). 
In \Fig{fig_oneloop},  we draw the one-loop self energy diagrams contributing to NLO. 
 Diagrams 
on the first line are one-loop self energy diagrams of the $G^{00}$ propagator and those on
the second line are for the $G^{zz}$ propagator. NLO corrections from the 
scalar field $\phi_I^a$  can only contribute to these self-energy diagrams.
The ``multi-gluon-exchange'' diagrams, such as those drawn in Fig.~\ref{fig_mul},
have no NLO corrections from the scalar fields and  
are thus equivalent to the QCD case calculated in \citep{CaronHuot:2008ni}.

Hence, diagrams (c), (i) and (j) in Fig.~\ref{fig_oneloop} are what we need to compute directly within ESYM. We
refer to App.~\ref{app_detail} for the detailed evaluation, whose final result reads
\begin{equation}
	\label{scalarcontrib}
\frac{\delta \cc_\mathrm{scalar}(q_\perp)}{g^4T^2C_R C_A}
=-\frac{3m_S}{2\pi(q_\perp^2+m_E^2)^2}+6\frac{m_S-\frac{q_\perp^2+4m_S^2}{2q_\perp}\tan^{-1}(q_\perp/2m_S)}{8\pi q_\perp^4},
\end{equation}
where the first term comes from diagram (c) and the second from (i) and (j).
This can be added to the QCD contribution in \cite{CaronHuot:2008ni} to give
\begin{align}
\frac{\delta\mathcal{C}(q_\perp)}{g^4T^2C_RC_A}&=\frac{7}{32q_\perp^3}+\frac{m_E}{4\pi(q_\perp^2+m_E^2)}\Big(\frac{3}{q_\perp^2+4m_E^2}-\frac{2}{q_\perp^2+m_E^2}-\frac{1}{q_\perp^2} \Big)\nonumber\\
&+\frac{-m_E-2\frac{q_\perp^2-m_E^2}{q_\perp}\tan^{-1}(\frac{q_\perp}{m_E})}{4\pi(q_\perp^2+m_E^2)^2}+\frac{m_E-\frac{q_\perp^2+4m_E^2}{2q_\perp}\tan^{-1}(\frac{q_\perp}{2m_E})}{8\pi q_\perp^4}\nonumber\\
&-\frac{\tan^{-1}(\frac{q_\perp}{m_E})}{2\pi q_\perp(q_\perp^2+m_E^2)}+\frac{\tan^{-1}(\frac{q_\perp}{2m_E})}{2\pi q_\perp^3}\nonumber\\
&-\frac{3m_S}{2\pi(q_\perp^2+m_E^2)^2}+6\frac{m_S-\frac{q_\perp^2+4m_S^2}{2q_\perp}\tan^{-1}(\frac{q_\perp}{2m_S})}{8\pi q_\perp^4}.
\label{nlocqperp}
\end{align}
This expression is well suited for the evaluation of $\qhat$ in Eqs.~\eqref{defqhat} and \eqref{splitqhat}. 
However, the
NLO collinear radiation rate we shall present in Sec.~\ref{sec_collinear} is most easily evaluated
from the impact-parameter space expression, Eq.~\eqref{defcxp}. To perform the needed Fourier transform,
we follow the same strategy used in \cite{Ghiglieri:2013gia}, that is to choose $\bxp=(\xp,0)$,
perform the integration for $q_y$ first and then for $q_x$, using contour techniques. When the latter cannot 
be done analytically, it has in any case become an integration with a (real) exponential
kernel rather than an oscillatory one, thus better suited for numerical integration.
Schematically, one has
\begin{equation}
\mathcal{C}'(\xp)\equiv\int\frac{d^2 q_\perp}{(2\pi)^2}(1-e^{i\bqp\cdot\bxp})\,
\mathcal{C}(q_\perp)
=\int^{\infty}_{-\infty}\frac{dq_x}{2\pi}(1-e^{i\xp q_x})\int^{\infty}_{-\infty}\frac{dq_y}{2\pi}\,\mathcal{C}(\qp)\,.
\end{equation}
At LO this becomes straightforwardly
\begin{equation}
	\mathcal{C}'^{(\text{LO})}(\xp)=\frac{g^2C_RT}{2\pi}(K_0(\xp m_E)+\gamma_E+\ln(\xp m_E/2))\,,
	\label{C_LO(b)}
\end{equation}
where this is intended to be the soft contribution, i.e. valid at distances $\xp\sim1/(gT)$.
At NLO we obtain $	\mathcal{C}'^{(\text{NLO})}(\xp)=	\mathcal{C}'^{(\text{LO})}(\xp)+\delta \cc'(\xp)$,
with the NLO correction given by
\begin{align}
&\frac{m_E\,\delta\mathcal{C}'(\xp )}{g^4T^2C_RC_A}\nonumber\\
=&\frac{3m_S}{8\pi^2m_E}(\xp m_E
K_1(\xp m_E)-1)-\frac{m_E}{16\pi^2m_S}\int^\infty_0\frac{dz}{z^4}(1-e^{-\xp m_Sz})(z^3-(z^2-4)^{3/2}\theta(z-2))\nonumber\\
&-\frac{1}{8\pi^2}\left\lbrace \frac{\xp m_EK_1(\xp m_E)+3-4e^{-\xp m_E}}{2}+\int^\infty_1 dz(e^{-\xp m_E}
-e^{-\xp m_Ez})\frac{\ln\frac{z^2}{z^2-1}}{(z^2-1)^{3/2}}\right\rbrace\nonumber\\
&-\frac{1}{96\pi^2}\int^\infty_0\frac{dz}{z^4}(1-e^{-\xp m_Ez})(z^3-(z^2-4)^{3/2}\theta(z-2))+\frac{7\xp m_E}{64\pi}-\frac{1}{32}\nonumber\\
&+\frac{1}{8\pi^2}\int^\infty_1 dz\frac{e^{-\xp m_Ez}\ln\frac{z^2}{z^2-1}}{\sqrt{z^2-1}}
+\frac{1}{8\pi^2}\int^\infty_0\frac{dz}{z^2}(1-e^{-\xp m_Ez})(z-\theta(z-2)\sqrt{z^2-4})\nonumber\\
&+\frac{1}{8\pi^2}(K_0(2\xp m_E)-2K_0(\xp m_E)+\ln\frac{4}{\xp m_E}-\gamma_E+\xp m_EK_1(\xp m_E)-1),
\label{C_NLO(b)}
\end{align}
where we have written the SYM-specific terms originating from soft scalar loops on the first line.  All 
other terms are shared with the QCD case in \cite{Ghiglieri:2013gia}.

\begin{figure}[ht]
\centering
\includegraphics[width=0.786\textwidth]{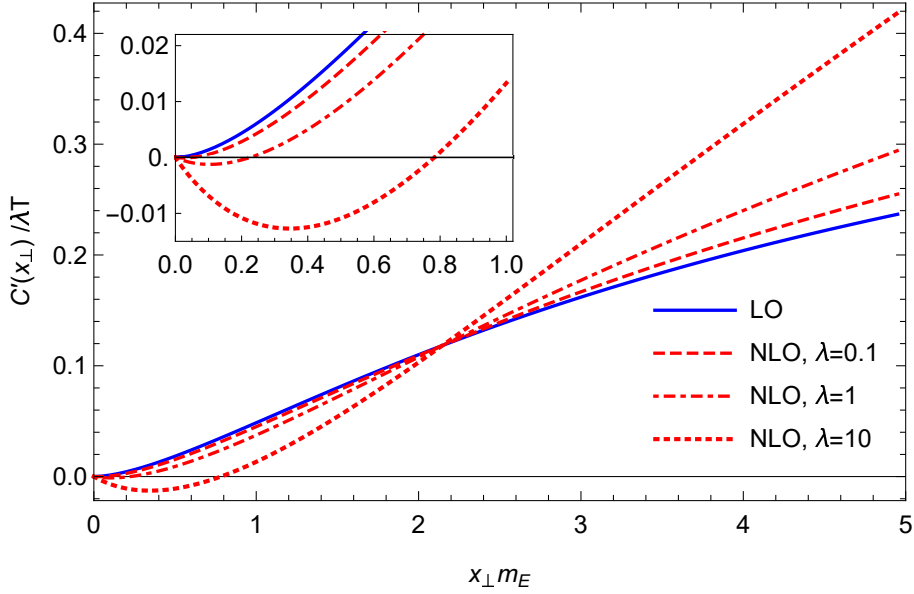}
\caption{$\mathcal{C}'(x_\perp)/\lambda T$ as a function of $x_\perp m_E$. The solid blue line
is the LO expression, Eq.~\eqref{C_LO(b)}, while the red lines are given by the sum of 
Eqs.~\eqref{C_LO(b)} and \eqref{C_NLO(b)}.}
\label{CLONLO}
\end{figure}
While we defer plots and considerations on the size of the NLO momentum-space corrections to the next section,
we plot in Fig.~\ref{CLONLO} the LO and NLO impact-parameter space kernels for three different
choices of coupling ranging from small to intermediate. As the plot shows, the NLO collision kernel becomes
negative close to the origin. That is quite obviously an UV ($\qp\gg gT$) effect: we will comment more on this 
in the next section. In the validity region of the soft calculation, $\xp m_E\sim 1$, the NLO curve can be either
below or above the LO one, while at the opposite asymptote at large $\xp$ it is larger.

\begin{figure}[ht]
\centering
\includegraphics[width=0.66\textwidth]{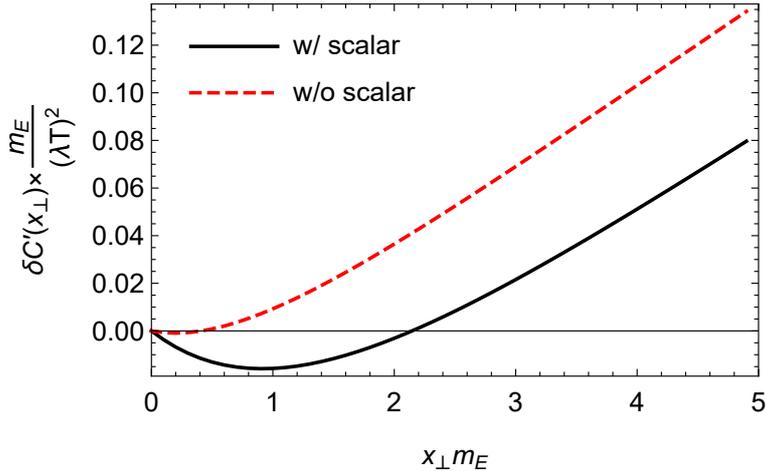}
\caption{$\delta\mathcal{C}'(x_\perp)\times m_E/(\lambda T)^2$ as a function of $x_\perp m_E$. The solid
black line is Eq.~\eqref{C_NLO(b)}, while the dashed red line is the QCD contribution only, without the
scalar loop contribution.}
\label{CNLOscalar}
\end{figure}
To show the impact of the new scalar contribution that is unique to SYM, in Fig.~\ref{CNLOscalar} we
plot the NLO correction $\delta\mathcal{C}'(x_\perp)$ with and without it, that is, in SYM and in QCD.
The figure clearly shows how the SYM scalar contribution magnifies
the negative dip already present in QCD, where it was noted in \cite{Ghiglieri:2013gia}. 
As we shall comment later, this negative dip will also have a marked impact on the collinear radiation rate.

\section{The transverse momentum broadening coefficient at NLO}
\label{sec_qhat}
As we illustrated in Sec.~\ref{ESYM}, the transverse momentum broadening coefficient $\hat{q}$ is given by 
the second moment of the collision kernel, as shown in Eq.~\eqref{defqhat}. We then introduced $q^*$  as
an intermediate regulator to separate the soft and hard contribution in Eq.~\eqref{splitqhat}. In the 
previous section we have reviewed the LO determination of $\cc_\mathrm{soft}(\qp)$ and computed its NLO
correction. The soft contribution
to $\qhat$ at LO and NLO  follow from Eqs.~\eqref{clo} and \eqref{nlocqperp} and are given by
\begin{align}
\label{loqhatsoft}
\hat{q}^{\text{(LO)}}_{\text{soft}}&=\frac{g^2TC_R m_E^2}{2\pi}\log\Big(\frac{q^*}{m_E}\Big),\\
\delta\hat{q}&=\frac{g^4T^2C_RC_A }{2\pi}\bigg[-\frac{q^*}{4}+m_E\frac{3\pi^2+10-4\log 2}{16\pi}+m_S\frac{3(1-\log 2)}{4\pi}\bigg],
\label{nloqhatsoft}
\end{align}
where
we have used the fact that $\log\frac{m_E}{m_S}=\frac{1}{2}\log2$ at leading order. The 
$m_S$-proportional contribution in Eq.~\eqref{nloqhatsoft} is the SYM-specific one arising from scalars. We
see that, while the LO contribution is logarithmically sensitive to $q^*$, as anticipated,
the NLO contribution presents a linear term in the cutoff and a finite contribution (where
$q^*$ has been taken to infinity in its evaluation). The linear term is related to the
negative dip in $\delta\cc'(\xp)$ at small $\xp$ observed previously: as we shall show later,
it will cancel against the hard contribution.

Eqs.~\eqref{loqhatsoft} and \eqref{nloqhatsoft} 
are already enough to determine $\qhat$ if $gT\ll q_\mathrm{max}\ll T$, as is the
case when dealing with diffusion processes in the Landau expansion of the collision operator
in the investigation of transport coefficients \cite{Ghiglieri:2018dib} or of high-energy partons
\cite{Ghiglieri:2015ala}. However, in the interest of generality, we have fixed $q_\mathrm{max}\gg T$,
as appropriate for the momentum broadening of a very hard parton, and we thus need to deal
with the hard contribution from the scale $T$, $\cc_\mathrm{hard}(\qp)$.

As we have mentioned before, the leading-order contribution comes
from elastic $2\leftrightarrow 2$ scatterings with medium constituents. In the soft
sector these get Landau-damped, leading to Eq.~\eqref{clo}, whereas in the
hard sector one can evaluate the matrix elements without any resummation and convolute
them with the statistical factors for the medium scatterers. In the QCD case this
gives \cite{Arnold:2008vd,CaronHuot:2008ni}
\begin{equation}
\mathcal{C}_{\text{hard}}^\mathrm{QCD}(q_{\perp})=\frac{g^4C_R}{q_\perp^4}\int\frac{d^3p}{(2\pi)^3}\frac{p-p_z}{p}
\Big[2C_A\,n_B(p)(1+n_B(p'))
+4 N_fT_F\,n_F(p)(1-n_F(p'))\Big],
\label{hardqcd}
\end{equation}
where $p'=p+\frac{q_\perp^2+2q_\perp\cdot p}{2(p-p_z)}$, $\nfd(p)=(e^{p/T}+1)^{-1}$ is the Fermi--Dirac distribution,
$T_F=1/2$ for the $N_f$ fundamental \emph{Dirac} fermions of QCD, 
not to be confused with $n_f=4$, which counts the number of \emph{Weyl} fermions of $\nfour$ SYM. We refer to \cite{Arnold:2008vd} for details on the accurate
numerical evaluation of this expression.

Eq.~\eqref{hardqcd} is transparently given by the sum of a contribution from scattering off gluons, proportional
to $C_An_B(1+n_B)$, and of a contribution from scattering off quarks and antiquarks, proportional to 
$N_fT_Fn_F(1-n_F)$. In principle the matrix elements for these scatterings, and for those involving external and 
intermediate scalars in $\nfour$ SYM, differ from each other (see \cite{Huot:2006ys,Czajka:2012gq} for a 
complete list). However, as noted in \cite{Arnold:2008vd,CaronHuot:2008ni,CaronHuot:2008uw}, 
when the external energy $E$ of the
very hard probe becomes much larger than $T$, the Mandelstam invariants $s\approx ET$ and $t=-\qp^2\sim T^2$
become hierarchically separated, with $s\gg |t|$. In this limit the matrix elements simplify greatly and acquire
a spin-independent, universal eikonal form proportional to $s^2/t^2$, which explains why the gluon and quark 
contributions in Eq.~\eqref{hardqcd} differ only in the statistical functions and group-theoretical factors.
In order to obtain the $\nfour$ SYM hard contribution we thus have to adjust the number and representation
of the scatterers in Eq.~\eqref{hardqcd}. In the bosonic sector, the $2C_A$ of QCD has to become $(2+6)C_A$,
to account for the two spin states of the gluon and the six real scalars. In the fermionic sector
$4 N_fT_F$ has to become $2n_f C_A$ to account for the $n_f$ Weyl fermions, which contribute each half as much as 
a Dirac fermion, and which transform in the adjoint representation, $T_A=C_A$. We then have
\begin{equation}
	\label{hardsym}
\mathcal{C}_{\text{hard}}^\mathrm{SYM}(q_{\perp})=\frac{g^4C_R}{q_\perp^4}\int\frac{d^3p}{(2\pi)^3}\frac{p-p_z}{p}
\Big[8C_A\,n_B(p)(1+n_B(p'))
+2n_fC_A\,n_F(p)(1-n_F(p'))\Big].
\end{equation}
The integration of Eq.~\eqref{hardsym} and its insertion
in Eq.~\eqref{splitqhat} give the hard contribution to $\qhat$ in $\nfour$ SYM, which reads
\begin{align}
\frac{\hat{q}_{\text{hard}}}{g^4C_RT^3}=&\frac{4C_A}{6\pi}\left[\log\left(\frac{T}{q^*}\right)+\frac{\zeta{(3})}{\zeta{(2)}}\log\left(\frac{q_\mathrm{max}}{T}\right)
-0.0688854926766592 \ldots +\frac{3}{16}\frac{q^*}{T} \right]\nonumber\\
&+\frac{n_fC_A}{12\pi}\left[\log\left(\frac{T}{q^*}\right)
+\frac{3}{2}\frac{\zeta{(3})}{\zeta{(2)}}\log\left(\frac{q_\mathrm{max}}{T}\right)-0.072856349715786
\ldots \right]+\mathcal{O}\left(\frac{q_*^2}{T^2}\right),
\label{qhathard}
\end{align}
where the high-precision numbers come from the QCD evaluation in \cite{Arnold:2008vd}.
As expected, this expression is logarithmically and linearly sensitive to the cutoff
$q^*$, canceling the opposite sensitivities in the soft sector in Eqs.~\eqref{loqhatsoft}
and \eqref{nloqhatsoft}. This corresponds to stating that the IR
limit of Eq.~\eqref{hardsym} is
\begin{equation}
	\label{hardsymlimit}
\mathcal{C}_{\text{hard}}^\mathrm{SYM}(q_{\perp}\ll T)=\frac{g^4C_RTm_E^2}{q_\perp^4}-\frac{g^4 C_R C_A T^2}{4\qp^3}
+\OO\left(\frac{g^4 T}{\qp^2}\right).
\end{equation}
As noted in \cite{Arnold:2008vd,CaronHuot:2008ni}, 
the IR linear sensitivity to the cutoff, caused by the second term above,
 emerges when taking the double limit
$\qp\to gT$ and $p\to gT$ in Eq.~\eqref{hardsym}. 
Because of the IR enhancement
of the Bose--Einstein distribution, only bosons cause this contribution. Since
in $\nfour$ SYM one has four times as many bosons to scatter from than in QCD
($2C_A$ gluons and $6 C_A$ scalars versus just $2C_A$ gluons), this contribution
is four times larger than in QCD. The cancellation of $q^*$-dependent
terms between Eqs.~\eqref{loqhatsoft}, \eqref{nloqhatsoft} and \eqref{qhathard} means
that, at the interface of the two regions, $gT\ll \qp\ll T$, the
soft and hard expressions for $\cc(\qp)$ must agree. This is trivially
verified for the first term in Eq.~\eqref{hardsymlimit} and Eq.~\eqref{clo}.
For what concerns the second term in Eq.~\eqref{hardsymlimit}, we
find that the UV limit of $\delta\cc(\qp)$ in Eq.~\eqref{nlocqperp}
is
\begin{equation}
	\label{uvlimit}
	\delta\cc(\qp\gg gT)=-\frac{g^4C_RC_AT^2}{4\qp^3}+\OO\left(\frac{1}{\qp^5}\right),
\end{equation}
which matches Eq.~\eqref{hardsymlimit}.
It is this term that causes the negative, linear dip at the origin in Fig.~\ref{CNLOscalar} (recall
that the dimensionally-regularized Fourier transform of $1/\qp^3$ in 2 dimensions is $-\xp/(2\pi)$).%
\footnote{The $1/\qp^3$ term was also extracted in the hard sector in the electroweak
theory in \cite{Ghiglieri:2016xye} (see Eq.~(D.6) there), finding $-g^2 C_R T^2/(16\qp^3)(C_A+N_S T_F)$, where $N_S=1$
is the complex, fundamental Higgs scalar doublet. Since $\nfour$ SYM has 6 \emph{real}, adjoint scalars, which implies
the replacement $N_ST_F\to n_sC_A/2=3C_A$, this 
expression is consistent with ours in Eq.~\eqref{uvlimit}.}

Summing Eqs.~\eqref{loqhatsoft}, \eqref{nloqhatsoft} and \eqref{qhathard} we obtain
the NLO  $\hat{q}$ of the $\nfour$ SYM plasma, which reads 
\begin{align}
\hat{q}=\frac{\lambda^2T^3}{6\pi}\bigg[&6\log\left(\frac{T}{m_E}\right)+7\frac{\zeta{(3})}{\zeta{(2)}}\log\left(\frac{q_\mathrm{max}}{T}\right)-0.4212546701382088\ldots\nn\\ 
&+\frac{m_E}{T}\left(\xi^{(\text{NLO})}_E+\xi^{(\text{NLO})}_S \right)\bigg]+\mathcal{O}(\lambda^3),
\label{nloqhat}
\end{align}
where the second line contains the NLO
contributions:  $\xi^{(\text{NLO})}_E=\frac{3}{16\pi}(3\pi^2+10-4\log 2)\simeq 2.198500$ is the
one also appearing in QCD, and $\xi^{(\text{NLO})}_S=\frac{9}{4\sqrt{2}\pi}(1-\log 2)\simeq 0.155399$ 
is the genuine SYM-specific contribution from soft scalars.
It is thus worth noting that this genuine SYM contribution has numerically a small impact, being less than
10\% of the gluon contribution shared with QCD. 
This should not be interpreted to mean that the overall
scalar contribution is small: recall that one half of $m_E$ and $m_S$ is due to (hard) scalars,
as shown in Eq.~\eqref{matchmasses}. In other words,
taking the QCD expression for $\delta\qhat$ and changing the Debye mass $m_E$
from the QCD to the SYM value, thus including the large hard scalar contribution thereto,
represents a good approximation to Eq.~\eqref{nloqhatsoft}. 
As we shall show in Sec.~\ref{sec_collinear},
 the overall large scalar contribution to $\delta\cc(\qp)$,
 in particular its limiting form in Eq.~\eqref{uvlimit}, has a significant impact on the NLO
 collinear splitting rate.

Let us quote the QCD results from \cite{Arnold:2008vd,CaronHuot:2008ni}
\begin{align}
\hat{q}_{\text{QCD}}=\frac{g^4C_RT^3}{6\pi}\bigg\{&
C_A\left[\log\left(\frac{T}{m_E}\right)
+\frac{\zeta{(3})}{\zeta{(2)}}\log\left(\frac{q_\mathrm{max}}{T}\right)
-0.0688854926766592 \ldots \right]\nonumber\\
+&N_fT_F\left[\log\left(\frac{T}{m_E}\right)
+\frac{3}{2}\frac{\zeta{(3})}{\zeta{(2)}}\log\left(\frac{q_\mathrm{max}}{T}\right)-0.072856349715786
\ldots \right]\nn\\
+&C_A\frac{m_E}{T}\xi^{(\mathrm{NLO})}_E\bigg\},
\label{qhatqcd}
\end{align}
where in QCD $m_E^2=g^2T^2(C_A+N_fT_F)/3$ and the term on the third line is the NLO contribution.
We note that in QCD the  NLO correction to $\qhat$ is large, as found in \cite{CaronHuot:2008ni}: already
at $\als=0.1$ it represent a 100\% correction: more precisely, for $N_f=3$,
$q_\mathrm{max}=50 T$ and $\als=0.1$ one finds $\qhat\approx 1.11 C_RT^3$ in LO QCD, $\qhat\approx 1.87 C_RT^3$ in 
NLO QCD. The very large $q_\mathrm{max}=50 T$ has been chosen to give a conservative
estimate of the NLO/LO ratio, as the LO contribution grows logarithmically
in $q_\mathrm{max}$: for a smaller $q_\mathrm{max}=10 T$ one would have
$\qhat\approx 0.59 C_RT^3$ in LO QCD, $\qhat\approx 1.35 C_RT^3$ in 
NLO QCD.
In the SYM case we have two observations: on the one hand,  
at LO in the hard region there is an extra, large contribution from scalars, which is exactly three times the gluon
contribution, together with a significant group-theory boost to the fermion contribution with respect to QCD,
while at NLO the numerical factor multiplying the expansion parameter $m_E/T$ changes by a small
amount with
respect to QCD, thus suggesting smaller relative NLO contributions.  
On the other hand, as we have mentioned, in $\nfour$ SYM $m_E/T$ is larger than in QCD at a given $g$,
due to the large scalar contribution thereto.
We then obtain, for $\als=0.1$, $N_c=3$ and $q_\mathrm{max}=50T$ we find in LO SYM $\qhat\approx10.20\,T^3$ and
in NLO SYM $\qhat\approx15.07\, T^3$ (for  $q_\mathrm{max}=10 T$ we have
$\qhat\approx 3.99\,T^3$ in LO SYM, $\qhat\approx 8.87\, T^3$ in 
NLO SYM).

Before we elaborate further on the NLO/LO ratio,
we must address the fact that, as pointed out in \cite{CaronHuot:2008ni}, Eqs.~\eqref{nloqhat} and \eqref{qhatqcd}
 suffer from truncation
effects arising from the intermediate regulator $q^*$. In
other words, Eqs.~\eqref{nloqhat} and \eqref{qhatqcd} have been obtained for $m_E\ll q*\ll T$ and can thus
become ill-behaved once $m_E\sim T$ ($\lambda\sim 1$).
 This is particularly important at leading order (i.e.
omitting the terms on the final line), where, for large enough $m_E\sim gT$, the entire expressions become negative.
Besides, the regulator makes a plot of $\cc(\qp)$ for all $\qp$ challenging. A solution to these two problems,
introduced in \cite{Arnold:2008vd} and used for the plots in \cite{CaronHuot:2008ni}, is to introduce
a new \emph{resummed scheme} for the LO $\cc(\qp)$, which reads (see also \cite{DEramo:2012uzl}
 where a different resummation is performed\footnote{In more detail,
 the authors of Ref.~\cite{DEramo:2012uzl} resum HTL self-energies for 
 exchanged momenta smaller than $q^*$, with $gT\ll q^*\ll T$, and full
 self-energies above that cut-off.})
\begin{equation}
	\label{defloscheme}
\mathrm{C}^{(\text{LO})}(q_\perp)\equiv \mathcal{C}_{\text{hard}}(q_\perp)\times\frac{q_\perp^2}{q_\perp^2+m_E^2}.
\end{equation}
At small $m_E/T$ (small $g$) it is easy to see that this expression, when plugged in Eq.~\eqref{defqhat},
reproduces the LO result for $\qhat$ plus some higher-order terms. In the IR, $\cc_{\text{hard}}(\qp\ll T)$
is dominated by the first term in Eq.~\eqref{hardsymlimit}
 and we thus reproduce Eq.~\eqref{clo}, the soft contribution to Eq.~\eqref{splitqhat}. In the hard region,
the factor $q_\perp^2/(q_\perp^2+m_E^2)$ becomes 1, up to corrections of order $m_E^2/\qp^2\sim g^2$. However,
due to these partially resummed higher-order corrections, the resulting curve extrapolates better to higher values
of $g$.

\begin{figure}[ht]
\centering
\includegraphics[scale=0.85]{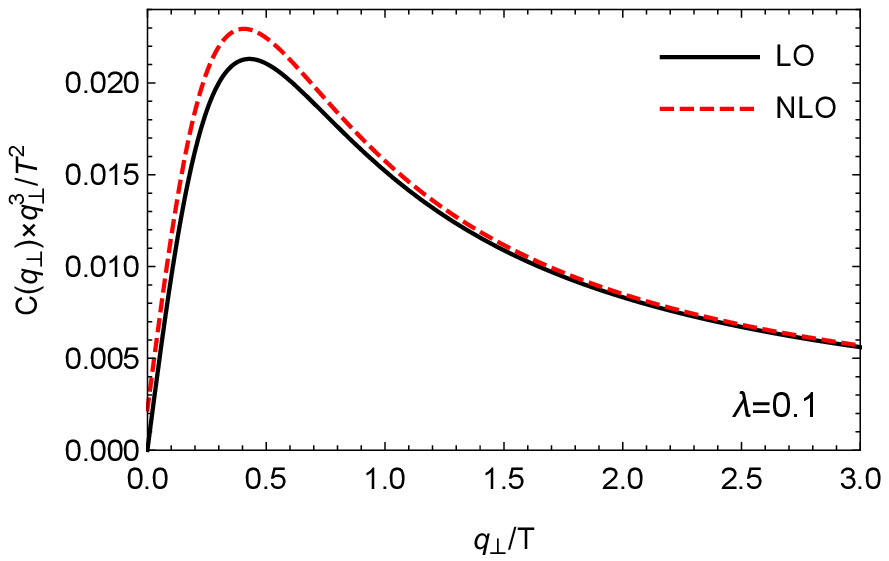}$\;$\includegraphics[scale=0.85]{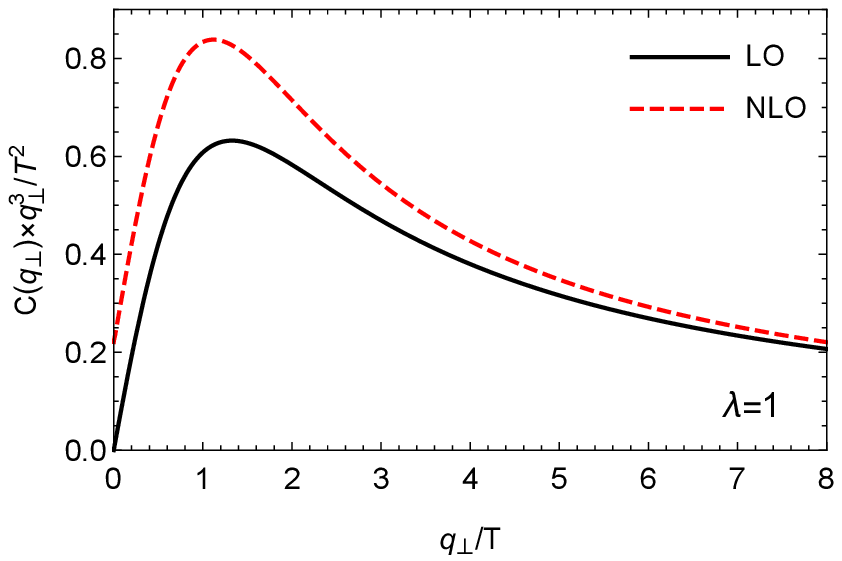}
\caption{$\mathrm{C}(q_\perp)\times q_{\perp}^3/T^2$ as a function of $q_{\perp}/T$ for different values of the 
coupling. The solid black lines come from Eq.~\eqref{defloscheme}, while the dashed red ones
from Eq.~\eqref{defnloscheme}. The area under the curves is directly proportional to $\qhat$.}
\label{CNLOscalar2}
\end{figure}
\begin{figure}[ht]
\centering
\includegraphics[scale=0.91]{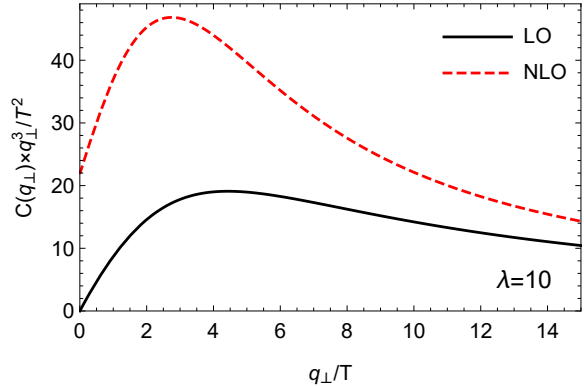}
\caption{$\mathrm{C}(q_\perp)\times q_{\perp}^3/T^2$ as in Fig.~\ref{CNLOscalar2}, at larger coupling.}
\label{CNLOscalar2large}
\end{figure}

At NLO one could simply add $\delta \cc(\qp)$ to this equation, which is the approach followed in
\cite{CaronHuot:2008ni}. However, this double-counts some NLO contributions, as $\cc_{\text{hard}}(\qp\ll T)$
contains also the subleading second term in Eq.~\eqref{hardsymlimit}
 giving rise to the linear-in-$q^*$ term in Eq.~\eqref{qhathard}, and which
is matched by the UV limit of $\delta\cc(\qp)$, Eq.~\eqref{uvlimit}. Hence, we propose instead the following
scheme for the NLO $\mathrm{C}(\qp)$
\begin{equation}
\mathrm{C}^{(\text{NLO})}(q_\perp)\equiv 
\mathcal{C}_{\mathrm{hard}}(q_\perp)+\mathcal{C}_\mathrm{soft}^{(\LO)}(q_\perp)
+\mathcal{C}_\mathrm{soft}^{(\text{NLO})}(q_\perp)
-\frac{2\lambda^2T^3}{\qp^4}+\frac{\lambda^2T^2}{4\qp^3}.
\label{defnloscheme}
\end{equation}
Contrary to the leading-order one, it is a \emph{strict} scheme: in the IR $\mathcal{C}_{\mathrm{hard}}(q_\perp)$ 
is approximated by the last two terms in this equation, as shown in
 Eq.~\eqref{hardsymlimit}. 
Hence
 $\mathcal{C}_{\mathrm{hard}}(q_\perp)$ cancels against these two terms there (up to terms of order $\frac{\lambda^2T}{\qp^2}$), leaving just $\mathcal{C}_\mathrm{soft}^{(\LO)}(q_\perp)
+\mathcal{C}_\mathrm{soft}^{(\text{NLO})}(q_\perp)$. In the UV, conversely, the last two terms cancel the UV
limits of the soft terms, leaving just $\mathcal{C}_{\mathrm{hard}}(q_\perp)$. It is easy to see that 
Eq.~\eqref{defnloscheme}, when inserted in Eq.~\eqref{defqhat}, reproduces Eq.~\eqref{nloqhat},\footnote{%
Up to discrepancies of order $1/q_\mathrm{max}$  from truncating the integration at $q_\mathrm{max}$,
whereas the finite terms $\xi^{(\mathrm{NLO})}$ in Eq.~\eqref{nloqhat} have been obtained with
$q_\mathrm{max}\to\infty$.} which further confirms that our Eq.~\eqref{nloqhat} is a strict prescription
for $\qhat$ at NLO. We plot our choices for $C(\qp)$ at LO and NLO in solid black
and dashed red in Figs.~\ref{CNLOscalar2} and \ref{CNLOscalar2large} 
for different values of the couplings
and multiplied by $\qp^3$, so that the area under these curves is directly proportional to $\qhat$. Similarly
to the QCD plots in \cite{CaronHuot:2008ni}, we see how the two curves differ more and more as the coupling
is increased. At $\lambda=10$, which is a typical ``intermediate'' coupling in heavy ion phenomenology, corresponding,
for $\nc=3$, to $\als\approx0.26$, the NLO corrections have completely overtaken the LO curve, signaling a convergence
problem of the perturbative expansion for this observable.

\begin{figure}
	\begin{center}
		\includegraphics[width=7.4cm]{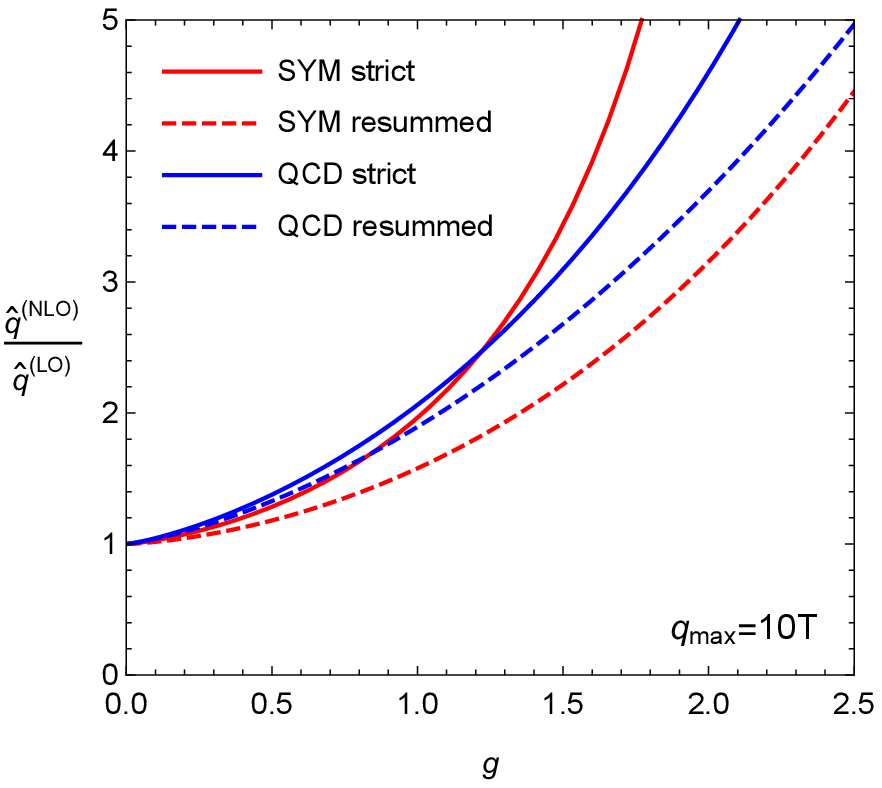}$\;$
		\includegraphics[width=7.4cm]{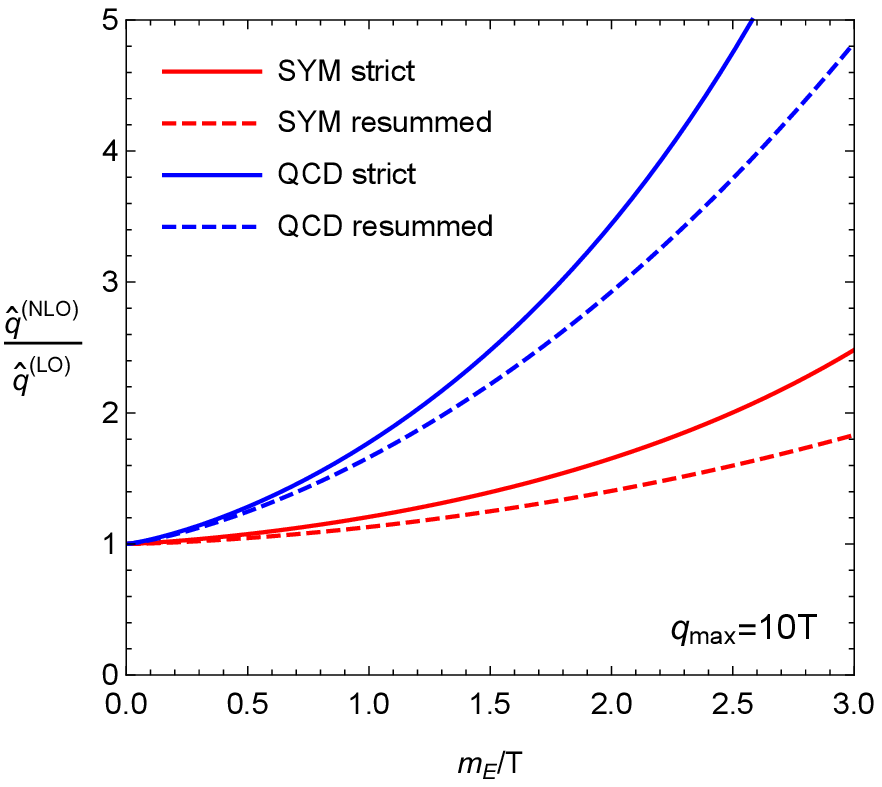}
	\end{center}
	\caption{The NLO/LO ratios for $\qhat$ in SYM and QCD. For
	the latter theory $\nc=N_f=3$, for the former $\nc=3$ on the left,
	while the r.h.s. is $\nc$-independent.
	$q_\mathrm{max}=10T$ in both plots. We refer to the text for the precise references
	to the equations being plotted.}
	\label{fig_ratios}
\end{figure}
In Fig.~\ref{fig_ratios} we plot instead the NLO/LO ratio for $\qhat$ in SYM and $\nc=N_f=3$ QCD. At NLO
we use the strict expression in Eqs.~\eqref{nloqhat} and \eqref{qhatqcd}. At LO, the
``strict'' curves use again those expressions omitting the NLO terms in their final lines.
The ``resummed'' curves are instead obtained by numerical integration of Eq.~\eqref{defloscheme},
with Eqs.~\eqref{hardqcd} and \eqref{hardsym} for QCD and SYM respectively. The plot on the l.h.s.
is as a function of $g$ (with $\nc=3$ for SYM as well) and shows that
NLO corrections are large in both theories at equal values of the coupling.
Furthermore, the size of the NLO corrections is \emph{similar} in the 
two theories and the discrepancy between the two schemes provides a first estimate of the truncation
uncertainty. On the right we plot the same equations as a function of $m_E$: as expected,
NLO corrections are significantly smaller in SYM at equal Debye mass.

Finally, we observe that Eq.~\eqref{defnloscheme} could also have applications in QCD where, to the best
of our knowledge, it has not appeared in the literature. In particular, we find that it is much better suited than
Eq.~\eqref{defloscheme} to a numerical Fourier transform. Eq.~\eqref{defloscheme} requires to sample two
numerical parameters, $m_E/T$ (or $g$) and $x_\perp T$, while the 
Fourier transforms of the individual terms in Eq.~\eqref{defnloscheme}
depend either on $x_\perp m_E$ or on $x_\perp T$, and many of them can be obtained analytically. We have tested
that the numerical Fourier transform of $\cc_\mathrm{hard}(\qp)$ is not particularly complicated once
a fine enough sampling of $\cc_\mathrm{hard}(\qp)$ has been obtained. This procedure leads to an 
expression for $\cc'(\xp)$ at NLO that interpolates smoothly from $\xp T\sim1$ to $\xp m_E\sim 1$ and would
thus be helpful for solutions of the collinear splitting equations 
in cases where
the energy of the mother parton is much larger than the temperature, requiring to account for transverse 
kicks at the scale $T$ beyond those at the scale $m_E$ normally included in the AMY formalism
\cite{Arnold:2002ja,Jeon:2003gi}.

\subsection{Comparisons with AdS/CFT  results}
\label{sec_ads}
Refs.~\cite{Liu:2006ug,Liu:2006he,D'Eramo:2010ak} computed the Wilson loop in Eq.~\eqref{defwloop}
for an adjoint source at strong coupling through the AdS/CFT correspondence. In its validity region, the result is 
Gaussian in $\xp$, i.e.
\begin{equation}
	\label{adsW}
	\vev{W(\xp,L)}_\mathrm{AdS}\approx\exp\left(-\frac{\pi^{3/2}\Gamma(3/4)}{4\Gamma(5/4)}\sqrt{\lambda} 
	T^3\,\xp^2 L\right).
\end{equation}
From our discussion in Sec.~\ref{ESYM} it follows that $\vev{W(\xp,L)}=\exp(-\qhat \xp^2 L/4)$ at small
$\xp$ and thus \cite{Liu:2006ug,Liu:2006he,D'Eramo:2010ak}
\begin{equation}
	\label{adsqhat}
	\qhat_\mathrm{AdS}=\frac{\pi^{3/2}\Gamma(3/4)}{\Gamma(5/4)}\sqrt{\lambda} 
	T^3\,.
\end{equation}
This result does not require a UV regulator: intuitively, the probability distribution $P(\qp)$,
which, as remarked in Footnote~\ref{foot_prob},
 is the Fourier transform of $\vev{W(\xp,L)}$ and which is related, but not equal, to 
$\cc(\qp)$ \cite{D'Eramo:2010ak,Benzke:2012sz,DEramo:2012uzl}, is also a Gaussian, with a finite
second moment, $\qhat$. In other words,
a conformal, strongly coupled description stays strongly coupled at all scales, while the need for 
a regulator at weak coupling arises from the $1/\qp^4$ UV tail of the $2\leftrightarrow 2$ scatterings,
i.e. rare large angle scatterings, oftentimes termed \emph{Moli\`ere scatterings} \cite{Moliere:1947zza,Moliere:1948zz}.
But such a quasi-particle picture cannot ever emerge in a strongly-coupled CFT,
in contrast with asymptotically-free QCD.

Furthermore, Eqs.~\eqref{adsW} and \eqref{adsqhat}, although obtained for $\lambda\to\infty$, depend 
on $\lambda$. This, together with the $q_\mathrm{max}$ dependence of the weak-coupling result,
make an attempt to extrapolate the weak- and strong-coupling results toward each other ill-posed
for $\qhat$. 
Observable quantities like the thermal photon rate or the shear viscosity,
which are $\lambda$-independent at strong coupling
and regulator-independent at weak coupling,
would make for much more sensible candidates for this type of comparison: our work in 
determining $\qhat$ and $\cc(\qp)$ at NLO represents an important stepping stone
towards NLO evaluation of these quantities.

However, there is still an important comparison that we can make and draw lessons from. Motivated
by the $\sqrt{\lambda}$ scaling in Eq.~\eqref{adsqhat}, Ref~\cite{Liu:2006he}
conjectured that, at strong coupling, $\qhat$ should scale like the square root of the entropy density $s$, i.e.
\begin{equation}
	\label{strongscaling}
	\frac{\qhat_\mathrm{QCD}}{\qhat_\mathrm{SYM}}\sim\sqrt{\frac{s_\mathrm{QCD}}{s_\mathrm{SYM}}}=\sqrt{\frac{47.5}{120}}
	\approx 0.63,
\end{equation}
where the entropy densities have been taken in the non-interacting limit for $N_c=3$ and $N_f=3$. 
The ratio does not change qualitatively
at stronger couplings, where the SYM entropy becomes (at large $N_c$, though) 
3/4 of the value above \cite{Gubser:1998nz}, as do to a 
good degree lattice QCD results in the transition region (see e.g. \cite{Borsanyi:2013bia,Bazavov:2014pvz}) as well.

\begin{figure}[ht]
	\begin{center}
		\includegraphics[width=10cm]{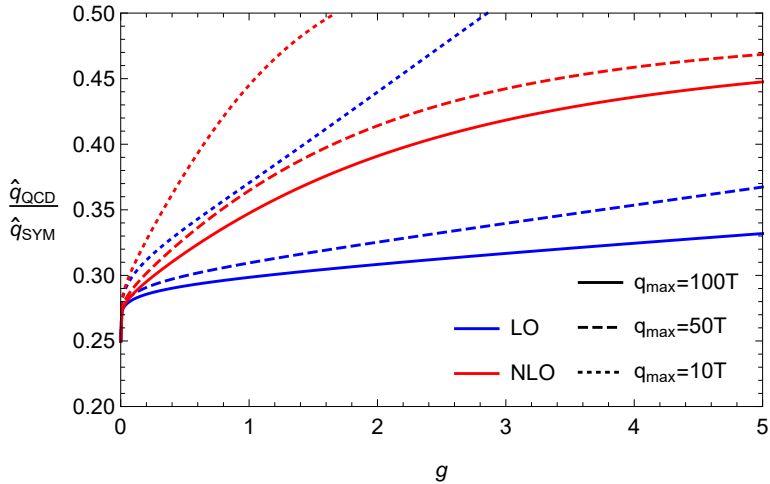}
	\end{center}
	\caption{The ratio of $\qhat$ for an adjoint QCD source to that in $\nfour$ SYM for three
	different values of $q_\mathrm{max}$ as a function of the coupling $g$, 
	at LO (blue) and NLO (red). The NLO SYM curves come from Eq.~\eqref{nloqhat} and the
	QCD ones from \eqref{qhatqcd}. At LO we instead integrate numerically Eq.~\eqref{defloscheme}
	in the two theories up to $q_\mathrm{max}$
	In both cases we have fixed $N_c=3$, in the case of QCD $N_f=3$. We truncate the curves for the smaller
	values of $q_\mathrm{max}$ shortly before the LO strict prescription
	would start to break down, which we take as an indicator of the point where the perturbative
	results are becoming unreliable, as the scale separation between $q_\mathrm{max}$, $T$
	and $m_E$ is disappearing (see also Fig.~\ref{fig_ratios}).
	}
	\label{fig_qhatratio}
\end{figure}
At weak coupling, we are now in the position of examining the QCD/SYM ratio at leading- and next-to-leading order,
as we do in Fig.~\ref{fig_qhatratio} as a function of the coupling for three different values of $q_\mathrm{max}$.
At NLO we choose our strict prescription, Eqs.~\eqref{nloqhat} and \eqref{qhatqcd},
while for LO we choose the resummed prescription in Eq.~\eqref{defloscheme}.
 In the QCD case we consider the $R=A$, i.e. take the $\qhat$
relevant for gluons.
As the plot shows, the curves start from 0.25, which
is the zero-coupling limit for this ratio for $N_c=3$ and $N_f=3$, and then grow to be
in the ballpark of 0.4, which does not differ too much from Eq.~\eqref{strongscaling}, even though,
as an inspection of Eqs.~\eqref{nloqhat} and \eqref{qhatqcd} shows, the dependence on
the number and type of degrees of freedom is not in the form of that of the square root
of the entropy density. We recall that the $\qhat$ relevant for quarks
is obtained through an extra factor of $C_F/C_A=4/9$. (Casimir scaling holds
to NLO, as Eq.~\eqref{qhatqcd} shows.)

In summary, if $q_\mathrm{max}$ in QCD is chosen at a value of a few times/10 times the
temperature, so as to encompass all the region that can be considered strongly coupled,
leaving the UV Moli\`ere tail to perturbation theory, then a recipe for using the AdS/CFT
results for $\qhat$ with a rescaling factor of order 1/2 would not be inconsistent
with our results for the QCD/SYM ratio at weak coupling. We also point out
that  in the region $\qp\lesssim E$, which is excluded from our  approach, as argued
in Sec.~\ref{ESYM}, the transferred momentum affects the kinematics of the hard parton
and the eikonal, Wilson-line based approach fails. We refer to \cite{DEramo:2018eoy}
for calculations of Moli\`ere scattering without the eikonal approximation.

\section{Collinear radiation rate}
\label{sec_collinear}
As we mentioned previously, $\cc(\qp)$ and $\qhat$ are important ingredients in the
determination of the medium-induced, collinear radiation rate. At weak coupling,
the leading-order photon and gluon radiation rates have been determined
in \cite{Arnold:2001ba,Arnold:2002ja} in QCD and extended to $\nfour$
SYM in \cite{CaronHuot:2006te,Huot:2006ys}. In QCD they have been 
extended to NLO in \cite{Ghiglieri:2013gia,Ghiglieri:2015ala}. As
a first application of our results of Sec.~\ref{sec_collisionkernel},
we now set out to extend the $\nfour$ SYM photon rate to NLO. The
extension to the gluon radiation rate is also straightforward, as it requires
the adaptation of the methodology used to extend the LO gluon radiation
rate to NLO to the extra scalar$\to$ scalar,gluon process of $\nfour$ SYM. 

The thermal photon production rate\footnote{
$\nfour$ SYM does not contain a photon. Ref.~\cite{CaronHuot:2006te}
gauged a $U(1)$ subgroup of the $R$-current to mimic electromagnetic
interactions in SYM. Two Weyl fermions
and two scalars become charged under this $U(1)$ interaction.} at leading order is given by \cite{CaronHuot:2006te}
\begin{align}
(2\pi)^3\frac{d\Gamma_\gamma}{d^3k}=\frac{1}{2k}g^{\mu\nu}W^<_{\mu\nu}(K),
\end{align}
where $k$ is the photon's momentum and
 $W^<_{\mu\nu}(K)$ is the backward Wightman two-point function of the $U(1)$ current
\begin{align}
W^<_{\mu\nu}(K)=\int d^4X e^{-iK\cdot X}\vev{J_\mu(0)J_\nu(X)}.
\end{align}
This rate receives LO contributions from Compton-like $2\leftrightarrow 2$ scatterings and from collinear
radiation, that is, collinear bremsstrahlung induced by the soft scatterings, governed by $\cc(\qp)$, from the charged, 
hard ($p\sim T$)
Weyl fermions and scalars, and its crossed process, collinear pair annihilation of the charged particles into the photon.
Collinearity ensues from the small momentum transfer from soft scatterings, which in turn causes an enhancement to these
rates, which would naively seem suppressed with respect to the $2\leftrightarrow 2$ component. Furthermore, collinear
emissions imply long formation times, which turn out to be of the same order of the inverse rate given by $\cc(\qp)$,
causing Landau-Pomeranchuk-Migdal (LPM) interference and requiring resummation. 

The LO collinear photon production rate, accounting for LPM resummation, is  given by \cite{CaronHuot:2006te}
\begin{align}
\gamma(k)&\equiv\frac{4\pi}{(N_c^2-1)g^2N_cT^2n_f(k)}g^{\mu\nu}W^<_{\mu\nu}(K)\Big|_{\text{coll}}\nonumber\\
&=\int^{\infty}_{-\infty} dp^+ \Big[\frac{n_F(k+p^+)(1-n_F(p^+))(p^{+2}+(k+p^+)^2)}{4n_F(k)p^{+2}(p^++k)^2}+\frac{n_B(k+p^+)(1+n_B(p^+))}{2n_F(k)p^+(k+p^+)}\Big]\nonumber\\
&\times \frac{1}{g^2N_cT^2}\int \frac{d^2p_\perp}{(2\pi)^2}\text{Re}[2\p_\perp\cdot \bm{f}(\p_\perp,p^+,k)],
\label{collinearrate_norm}
\end{align}
where the normalization of $\gamma(k)$ has been chosen as the leading-log coefficient of the $2\leftrightarrow 2$
component, which we do not consider here.  The photon momentum $k$ has been chosen along the $z$ direction.
The $n_F$- and $n_B$-proportional terms are the contribution from bremsstrahlung from (pair annihilation of)
fermions and scalars respectively.
The function $\bm{f}(\p_\perp,p^+,k)$ is the solution of the following integral equation which
resums an infinite number of soft scatterings, thus accounting for LPM interference
\begin{align}
2\p_\perp&=\frac{ik(p_\perp^2+m_\infty^2)}{2p^+(k+p^+)} \bm{f}(\p_\perp,p^+,k)+\int \frac{d^2q_\perp}{(2\pi)^2}\mathcal{C}(q_\perp)[\bm{f}(\p_\perp,p^+,k)-\bm{f}(\p_\perp+\q_\perp,p^+,k)].
\label{integral_eq}
\end{align}
The first term on the r.h.s. is a kinetic term: 
it encodes the energy difference between the final and initial states,
caused by the soft scatterings described by second term, the collision operator.
$m_\infty$ in the kinetic term is the thermal asymptotic mass of the emitter fermions and scalars, which
for $p\sim T$ obey the dispersion relation $p_0=\pm\sqrt{p^2+m_\infty^2}$. In $\nfour$ SYM these
asymptotic masses preserve supersymmetry, as they are the same for all species (gluons, fermions and scalars):
$m_\infty^2=m_E^2/2=m_S^2=\lambda T^2$ at leading order \cite{CaronHuot:2008uw}.

NLO $\OO(g)$ corrections to Eq.~\eqref{collinearrate_norm} can enter only in the two inputs
that are sensitive to the $gT$ scale, as proven in \cite{Ghiglieri:2013gia}. These are $\cc(\qp)$,
which we have just computed to NLO, and $m_\infty$, whose $\OO(g)$ correction has been computed
in \cite{CaronHuot:2008uw}, finding $\delta m^{2}_\infty=-g^3C_A^{3/2}T^2(3+\sqrt{2})/(2\pi)$.
Hence, to find the NLO corrections to $\gamma(k)$, $\bm{f}$ can be treated as an expansion
in powers of $\delta m_\infty^2$ and $\delta\cc$. The zeroth-order reproduces the LO
expression, and the first order in each of the two corrections gives $\gamma^{\delta m}$
and $\gamma^{\delta \mathcal{C}}$, so that 
$\gamma^{\text{(NLO)}}=\gamma^{\text{(LO)}}+\gamma^{\delta m}+\gamma^{\delta \mathcal{C}}$.
In order to determine numerically these functions, it is convenient
to Fourier-transform Eq.~\eqref{integral_eq} in $x_\perp$-space, where
it becomes a two-dimensional Schr\"odinger-like equation with an imaginary potential
given by $\cc'(\xp)$ \cite{Aurenche:2002wq}. We refer to  \cite{Ghiglieri:2013gia} for details
on the strategy to solve Eq.~\eqref{integral_eq} at NLO.

\begin{figure}[ht]
\centering
\includegraphics[scale=1]{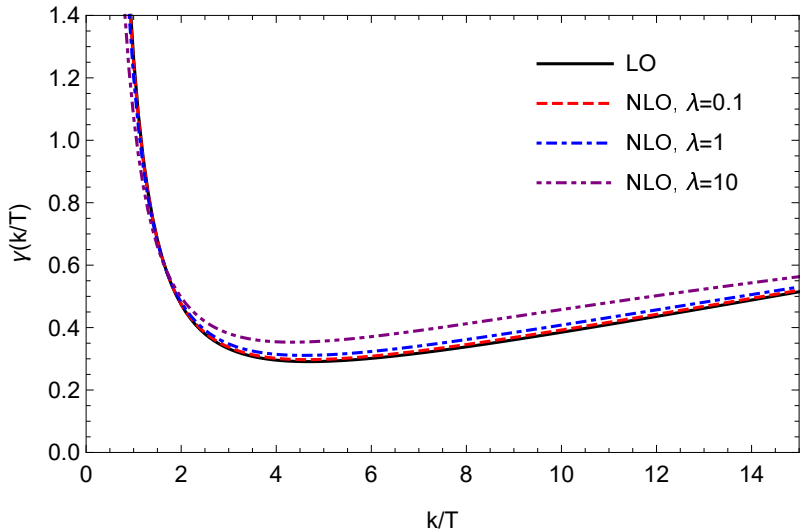}
\caption{The collinear photon  rate in $\nfour$ SYM. We plot the LO results \cite{CaronHuot:2006te}
in solid black and the NLO results at different couplings in various colors and dashing patterns.}
\label{plot_collinear}
\end{figure}
In Fig.~\ref{plot_collinear} we plot our NLO results for $\gamma$. As the figure clearly show,
the NLO corrections turn out to be  small: even at $\lambda=10$ they represent at most a 30\%
decrease in the IR and a 20\% increase at $k\approx 5T$. This is in sharp contrast with
the results in NLO QCD \cite{Ghiglieri:2013gia}, where for $\als=0.3$ the correction is approximately 
a 100\% increase.

\begin{figure}[ht]
\centering
\includegraphics[scale=0.7]{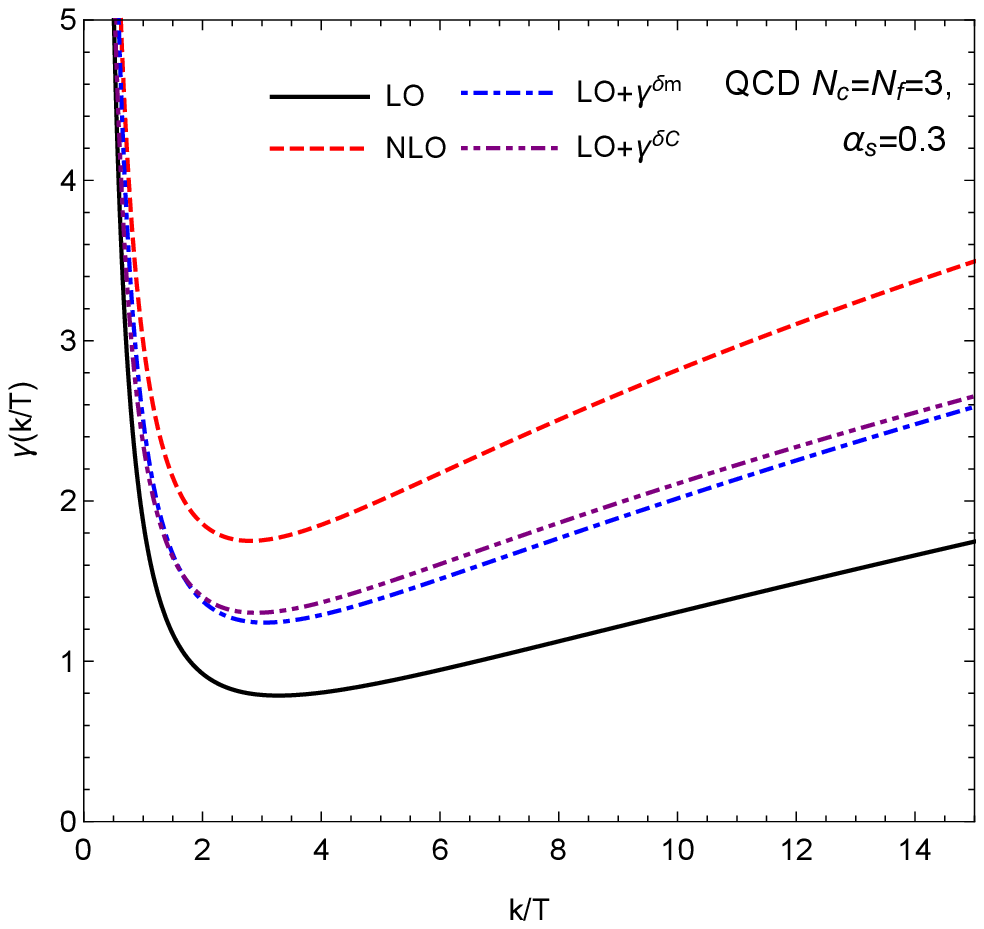}$\;$
\includegraphics[scale=0.715]{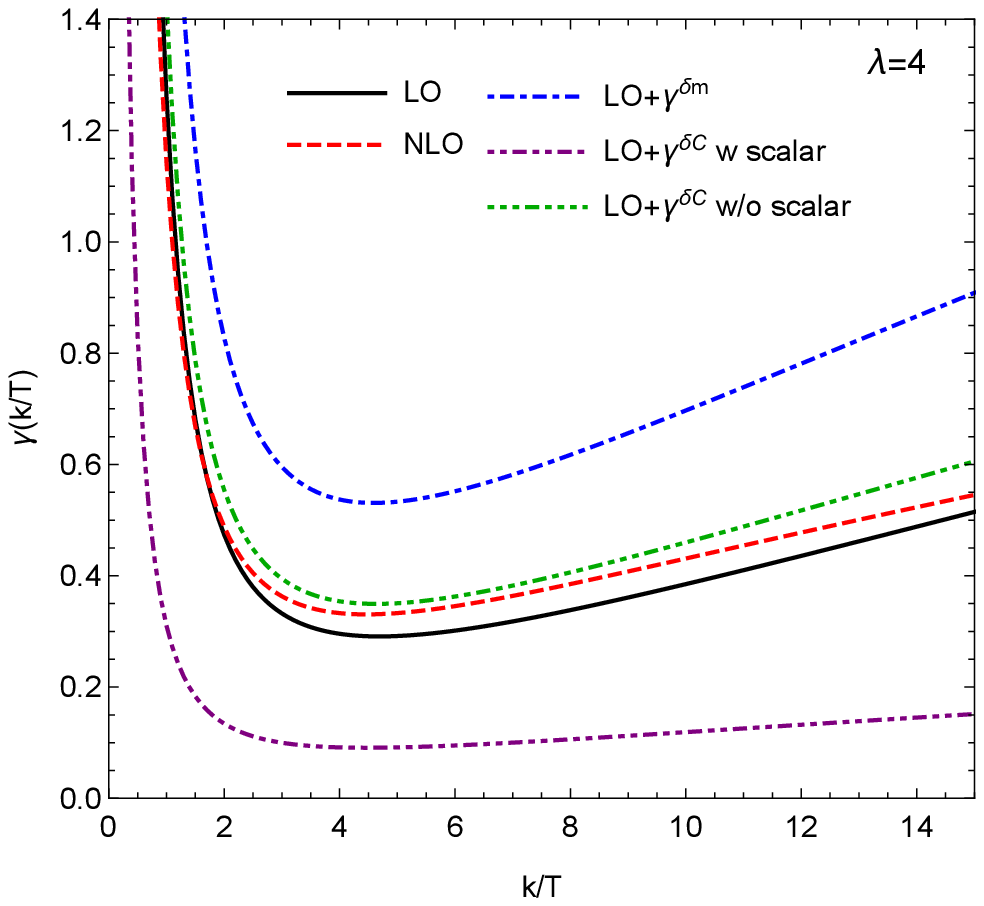}
\caption{Left: the separate sum of the LO $\gamma$ with each  of the two NLO
corrections
 $\gamma^{\delta m}$ and $\gamma^{\delta \mathcal{C}}$ in QCD
 with $\nc=N_f=3$, $\als=0.3$. The results
 are taken from \cite{Ghiglieri:2013gia} and are normalized by the QCD leading-log coefficient.
 On the right we plot the same in $\nfour$ SYM for $\lambda=4$, with the addition of the extra
 curve for $\gamma^{\delta\cc}$ where $\delta\cc$ is taken without the scalar 
 contribution~\eqref{scalarcontrib},
 corresponding to the QCD $\delta\cc$.
In both cases the full NLO collinear emission rate, $\gamma^{\text{(NLO)}}$, is a 
sum of the two corrections $\gamma^{\delta m}$
 and $\gamma^{\delta \mathcal{C}}$ and of the leading order result $\gamma^{\text{(LO)}}$. }
\label{plot_collinearscalar}
\end{figure}
To investigate this large discrepancy, let us look at Fig.~\ref{plot_collinearscalar}. 
On the left we plot the QCD results of \cite{Ghiglieri:2013gia} (with $\nc=N_f=3$, normalized by the QCD leading-log coefficient), split in
the contribution of $\delta m_\infty^2$ and $\delta\cc$ for $\als=0.3$. On the right
we plot the same for SYM. We see that the $\gamma^{\delta m}$ curves have a very similar behavior,
but the $\gamma^{\delta \cc}$ do not. In fact, the latter is positive and comparable with $\gamma^{\delta m}$
in QCD, while it is negative in SYM. These facts can be understood as follows:
the $\gamma^{\delta m}$ correction is proportional to $\delta m_\infty^2/m_\infty^2$. In QCD
this is $-2m_E/(\pi T)$, which, for $\nc=N_f=3$, is approximately $-0.78g$. In SYM
it approximates $-0.70\sqrt{\lambda}$, so for our choice of $\lambda=4$ and $\als=0.3$,
for which $g\approx\sqrt{\lambda}=2$, we expect the observed similar behavior.
For the $\delta\cc$ correction, on the other hand, there are two major factors to account for.
As the SYM plot shows, if we omit the scalar contribution, Eq.~\eqref{scalarcontrib},
from $\delta\cc$, corresponding to using
the $\delta\cc$ of QCD,  the resulting correction becomes positive, as in QCD. And as
the discussion in Secs.~\ref{sec_collisionkernel} and \ref{sec_qhat}, together with Fig.~\ref{CNLOscalar},
illustrated, the scalar contribution has the effect of making $\delta \cc'(\xp)$ smaller, accentuating
its negative dip at small $\xp$,
due to the large UV, $\lambda^2/\qp^3$-proportional term which is three times the one in the gluon
contribution. Simple positivity argument imply that
it is the negative part of $\delta\cc'(\xp)$ that causes the observed negative $\gamma^{\delta\cc}$. Furthermore, $\gamma^{\delta\cc}$ is proportional to $g^2 C_AT/m_E$, which
is the ratio of the coefficients of $\delta\cc'(\xp)$ and $\cc'(\xp)$. In $\nc=N_f=3$ QCD this approximates to 
$2.45 g$, while in SYM it is the much smaller $0.71 \sqrt{\lambda}$, which
goes to explain why the curve without the scalar contribution to $\cc$, that is with the QCD $\delta \cc$,
has a much smaller impact than in QCD.

\section{Summary and conclusions}
\label{sec_concl}

In this paper we have investigated the transverse scattering kernel
$\cc(\qp)$ and the transverse momentum broadening coefficient $\qhat$ at NLO
in perturbative $\nfour$ SYM, including thus $\OO(g)$
effects from soft bosons. The NLO correction to the former, $\delta\cc(\qp)$,
has been evaluated in Sec.~\ref{sec_collisionkernel} using the analytical mapping
to a dimensionally-reduced Euclidean theory introduced in \cite{CaronHuot:2008ni}.
Our analysis in that section identified all diagrams with contributions
from the scalar fields of $\nfour$ SYM, that are absent in the QCD
calculation \cite{CaronHuot:2008ni} and that we evaluated, yielding our
result for the NLO correction to $\cc$, Eq.~\eqref{nlocqperp}. We also obtained
the Fourier-transformed, impact-parameter space expression in Eq.~\eqref{C_NLO(b)},
which is more useful in the solution of the collinear rate equation
and which shows how the NLO correction is negative at small $x_\perp$ and how the
scalar contribution boosts this feature, as shown
in Figs.~\ref{CLONLO} and \ref{CNLOscalar}.

In Sec.~\ref{sec_qhat} we used the NLO expression for the scattering kernel
to determine $\qhat$ at NLO, which may be found in Eq.~\eqref{nloqhat}.
This required also obtaining the contribution from the scale $T$ to $\qhat$
and understanding how the soft and hard contributions match in the interface
region between the two scales. In particular, we found that the hard contribution
from the SYM scalars is three times that from gluons, also present in QCD.
The NLO contribution is proportional to the Debye mass $m_E$, which
is also boosted by a large hard scalar contribution in a threefold ratio
with the gluon one. However, the  soft scalar loops we computed
 give a numerically small contribution, of the order of $10\%$ of the 
 soft gauge loops already computed in QCD.
 
We then used our NLO results to try to understand how to best bridge
between QCD and SYM. The strong-coupling AdS/CFT calculations 
\cite{Liu:2006ug,Liu:2006he,D'Eramo:2010ak} are not directly comparable,
because of the distinctive UV-divergences that arise at weak coupling
in $\qhat$ from large-angle scatterings. However, if one considers
$\qhat$ as an effective parameter describing scatterings with the medium
up to some $q_\mathrm{max}$ below which a strong-coupling
description is considered applicable, then a prescription for extrapolating
the AdS/CFT results to the different degrees of freedom  of QCD is necessary. 
Ref.~\cite{Liu:2006he} conjectured that a rescaling by the  square root of the ratio 
of the entropy densities in the two theories would provide a good measure. 
We have examined the QCD/SYM ratio for $\qhat$ at weak coupling at NLO,
see Fig.~\ref{fig_qhatratio}, and
found that it ranges from being half of the conjectured ratio at small couplings
to 2/3 of it once intermediate couplings are approached, thus potentially
indicative of an extrapolation to the conjectured strong-coupling ratio.
Furthermore, our weak-coupling ratio was not widely different between leading-
and next-to-leading order. In other words, the NLO/LO ratio is similar
in QCD and SYM for $\qhat$ at fixed coupling, as shown explicitly by our Fig.~\ref{fig_ratios}.

Our examination of the collinear splitting rate in Sec.~\ref{sec_collinear}
may however suggest that no universal lessons can be drawn from these particular NLO
corrections. Indeed, we found that the NLO collinear splitting rate
is a small ($\OO(10-20\%)$) modification of LO in SYM even at $\lambda=10$, see 
Fig.~\ref{plot_collinear}, in sharp contrast to
QCD where it is essentially twice the LO rate for $\als=0.3$. The reason is that these
corrections are very sensitive to the precise nature of the theory and its degrees of freedom.
In QCD both the $\OO(g)$ shift in the dispersion relation and the $\OO(g)$ shift in $\cc(\qp)$ 
result in positive corrections to the collinear rate. The weights of these corrections
are different and depend in a non-trivial way on the statistics and group theory properties
of the d.o.f.s: for $\nc=N_f=3$ QCD the corrections are almost identical (see Fig.~\ref{plot_collinearscalar}).
In SYM, on the other hand, these weights are in a very different ratio and furthermore
the correction due to $\delta\cc(\qp)$  is negative, for reasons we can
attribute to the large scalar contribution to $\cc(\qp)$ at the interface between the soft
and hard regions. Hence the two contributions experience a large cancellation,
which is reminiscent of the large, accidental cancellations
found between different NLO contributions in the QCD photon rate \cite{Ghiglieri:2013gia},
which was observed to be dependent on the specifics of the d.o.f.s. We thus conclude
that the cancellation we observe is also largely accidental and not motivated by any particular
symmetry such as supersymmetry: QCD with scalar quarks in place of fermion ones would also
be susceptible to such a cancellation, its precise amount sensitive to how many scalar quarks
are introduced and the representation they transform under.

Indeed, it would be interesting to see how the SYM photon rate is modified at NLO.
Besides these collinear corrections we have computed, one would need the scalar contribution
to the other kinematical regions identified in the QCD calculation, i.e. the semi-collinear
and soft ones. While the former seems rather straightforward, a preliminary analysis of the latter
shows that it would require a non-trivial calculation using the light-cone sum-rules of
\cite{Ghiglieri:2013gia,Ghiglieri:2015ala}. Similar work would be required also
for  an NLO computation of the shear
viscosity and other transport coefficients. It would be very interesting to understand whether
the large corrections found in \cite{Ghiglieri:2018dib} for QCD -- and almost entirely driven by its
large  $\delta\qhat$ -- 
also appear in the SYM case, with the comparable $\delta\qhat$ we have found.
Indeed, a major unanswered point raised by \cite{Ghiglieri:2018dib} is the identification
of the physics responsible for these large corrections  and the subsequent reorganization
of the perturbative series. 

To this end, one of the possible pathways is the lattice determination of the $gT$-scale physics. 
The mapping to the Euclidean theory we have exploited for our computation
in Sec.~\ref{sec_collisionkernel} allows also for lattice-EQCD determinations of $\cc(\qp)$
and related observables whose soft contribution is dominated by the Matsubara zero mode. Indeed,
a first study of $\qhat$ and $\cc'(\xp)$ in lattice EQCD has been carried out \cite{Panero:2013pla}
and could be extended to ESYM without encountering any of the major issues that
affect lattice discretizations of $\nfour$ SYM (see
\cite{Catterall:2009it,Bergner:2016sbv} for reviews), as supersymmetry is 
broken in ESYM and fermions are absent. The conformal nature of $\nfour$ SYM may further
simplify the matching to ESYM, potentially making (E)SYM a good testbed of
these lattice techniques in a program, featuring our results, of more precise measurements and 
investigations of other observables.

\section*{Acknowledgments}
The authors would like to thank  
Urs Wiedemann and Krishna Rajagopal for useful conversations. 
This work was supported by the Korean Research Foundation (KRF) through
the CERN-Korea graduate student program and was partially supported by the 
Graduate School of YONSEI University Research Scholarship Grants in 2017.

\appendix
\section{Feynman Rules of ESYM}
\label{app_feyn}
The Feynman rules are obtained from the ESYM Lagrangian,
Eq.~\eqref{esymlag}, in a  rather straightforward way. We collect them here
in Feynman gauge, together with our graphical notation. The propagators read
\begin{align}
\raisebox{-6pt}{
\begin{fmffile}{gauge}
\fmfset{curly_len}{2.0mm}
\begin{fmfgraph*}(40,20)\fmfpen{0.5thin}
\fmfleft{i1}
\fmfright{o1}
\fmflabel{$i;a$}{i1}
\fmflabel{$j;b$}{o1}
\fmf{gluon,label=$p$}{i1,o1}
\end{fmfgraph*}\end{fmffile}} &\qquad=\frac{\delta^{ab}\delta_{ij}}{p^2}\qquad:\;gauge\,,
\\
\raisebox{-8pt}{\begin{fmffile}{electrostatic}
\begin{fmfgraph*}(40,20)\fmfpen{0.5thin}
\fmfleft{i1}
\fmfright{o1}
\fmflabel{$0;a$}{i1}
\fmflabel{$0;b$}{o1}
\fmf{vanilla,label=$p$}{i1,o1}
\end{fmfgraph*}\end{fmffile}}
&\qquad=\frac{\delta^{ab}}{p^2+m_E^2}\;\;:\;electrostatic\,,\\
\raisebox{-8pt}{\begin{fmffile}{scalar}
\begin{fmfgraph*}(40,20)\fmfpen{0.5thin}
\fmfleft{i1}
\fmfright{o1}
\fmflabel{$I;a$}{i1}
\fmflabel{$J;b$}{o1}
\fmf{dashes,label=$p$}{i1,o1}
\end{fmfgraph*}\end{fmffile}}&\qquad=\frac{\delta^{ab}\delta_{IJ}}{p^2+m_S^2}\;\;:\;scalar\,,\\
\raisebox{-8pt}{\begin{fmffile}{ghost}
\begin{fmfgraph*}(40,20)\fmfpen{0.5thin}
\fmfset{arrow_len}{2mm}
\fmfset{dot_len}{1.5mm}
\fmfleft{i1}
\fmfright{o1}
\fmflabel{$a$}{i1}
\fmflabel{$b$}{o1}
\fmf{dots_arrow,label=$p$}{i1,o1}
\end{fmfgraph*}\end{fmffile}}&\qquad=\frac{\delta^{ab}}{p^2}\quad\quad\;\;\;:\;ghost\,.
\end{align}
The three-point vertices are
\begin{align}
\raisebox{-18pt}{\begin{fmffile}{3gauge}
\fmfset{curly_len}{2.0mm}
\begin{fmfgraph*}(40,40)\fmfpen{0.5thin}
\fmfleft{i1}
\fmfright{o1,o2}
\fmflabel{$i;a$}{i1}
\fmfv{label=\hspace{0.8em}$j;b$,label.dist=-2mm}{o1}
\fmfv{label=\hspace{0.8em}$k;c$,label.dist=-2mm}{o2}
\fmf{gluon,label=$p$}{i1,v1}
\fmf{gluon,label=$q$}{o1,v1}
\fmf{gluon,label=$r$,label.side=right}{o2,v1}
\end{fmfgraph*}
\end{fmffile}}&\quad=ig_Ef^{abc}(\delta_{ij}(p-q)_k+\delta_{jk}(q-r)_i+\delta_{ki}(r-p)_j),\\\nonumber\\
\raisebox{-18pt}{\begin{fmffile}{1gauge2static}
\fmfset{curly_len}{2.0mm}
\begin{fmfgraph*}(40,40)\fmfpen{0.5thin}
\fmfleft{i1}
\fmfright{o1,o2}
\fmflabel{$i;a$}{i1}
\fmfv{label=\hspace{0.8em}$0;b$,label.dist=-2mm}{o1}
\fmfv{label=\hspace{0.8em}$0;c$,label.dist=-2mm}{o2}
\fmf{gluon}{i1,v1}
\fmf{vanilla}{o1,v1}
\fmf{vanilla}{o2,v1}
\end{fmfgraph*}
\end{fmffile}}&\quad=ig_Ef^{abc}(q-r)_i,\\\nonumber\\
\raisebox{-18pt}{\begin{fmffile}{1gauge2scalar}
\fmfset{curly_len}{2.0mm}
\begin{fmfgraph*}(40,40)\fmfpen{0.5thin}
\fmfleft{i1}
\fmfright{o1,o2}
\fmflabel{$i;a$}{i1}
\fmfv{label=\hspace{0.8em}$I;b$,label.dist=-2mm}{o1}
\fmfv{label=\hspace{0.8em}$J;c$,label.dist=-2mm}{o2}
\fmf{gluon}{i1,v1}
\fmf{dashes}{o1,v1}
\fmf{dashes}{o2,v1}
\end{fmfgraph*}
\end{fmffile}}&\quad=ig_E\delta_{IJ}f^{abc}(q-r)_i,\\\nonumber\\
\raisebox{-18pt}{\begin{fmffile}{1gauge2ghost}
\fmfset{curly_len}{2.0mm}
\fmfset{dot_len}{1.5mm}
\fmfset{arrow_len}{2mm}
\begin{fmfgraph*}(40,40)\fmfpen{0.5thin}
\fmfleft{i1}
\fmfright{o1,o2}
\fmflabel{$i;a$}{i1}
\fmfv{label=\hspace{0.8em}$b$,label.dist=-2mm}{o1}
\fmfv{label=\hspace{0.8em}$c$,label.dist=-2mm}{o2}
\fmf{gluon}{i1,v1}
\fmf{dots_arrow}{o1,v1}
\fmf{dots_arrow}{v1,o2}
\end{fmfgraph*}
\end{fmffile}}&\quad=-ig_Ef^{abc}r_i.
\end{align}
Finally, the four-point vertices read
\begin{align}
\raisebox{-10pt}{\begin{fmffile}{4gauge}
\fmfset{curly_len}{2.0mm}
\begin{fmfgraph*}(45,35)\fmfpen{0.5thin}
\fmfleft{i1,i2}
\fmfright{o1,o2}
\fmfv{label=$k;c$,label.dist=-0.3mm}{i1}
\fmfv{label=$i;a$,label.dist=-0.3mm,label.angle=180}{i2}
\fmfv{label=$l;d$,label.dist=-0.3mm}{o1}
\fmfv{label=$j;b$,label.dist=-0.3mm,label.angle=0}{o2}
\fmf{gluon}{i1,v1}
\fmf{gluon}{i2,v1}
\fmf{gluon}{o1,v1}
\fmf{gluon}{o2,v1}
\end{fmfgraph*}
\end{fmffile}}&
\begin{array}{ll}
\quad=-g_E^2&[f^{abc}f^{cde}(\delta_{ik}\delta_{jl}-\delta_{il}\delta_{jk})+f^{ace}f^{bde}(\delta_{ij}\delta_{kl}-\delta_{il}\delta_{jk})\\
&+f^{adc}f^{bce}(\delta_{ij}\delta_{kl}-\delta_{ik}\delta_{jl})],
\end{array}\\\nonumber\\
\raisebox{-15pt}{\begin{fmffile}{2gauge2static}
\fmfset{curly_len}{2.0mm}
\begin{fmfgraph*}(45,35)\fmfpen{0.5thin}
\fmfleft{i1,i2}
\fmfright{o1,o2}
\fmfv{label=$0;c$,label.dist=-0.3mm}{i1}
\fmfv{label=$i;a$,label.dist=-0.3mm}{i2}
\fmfv{label=$0;d$,label.dist=-0.3mm}{o1}
\fmfv{label=$j;b$,label.dist=-0.3mm}{o2}
\fmf{vanilla}{i1,v1}
\fmf{gluon}{i2,v1}
\fmf{vanilla}{o1,v1}
\fmf{gluon}{o2,v1}
\end{fmfgraph*}
\end{fmffile}}&\quad=-g_E^2\delta_{ij}(f^{ace}f^{bde}+f^{bce}f^{ade}),\\\nonumber\\
\raisebox{-15pt}{\begin{fmffile}{2gauge2scalar}
\fmfset{curly_len}{2.0mm}
\begin{fmfgraph*}(45,35)\fmfpen{0.5thin}
\fmfleft{i1,i2}
\fmfright{o1,o2}
\fmfv{label=$I;c$,label.dist=-0.3mm}{i1}
\fmfv{label=$i;a$,label.dist=-0.3mm}{i2}
\fmfv{label=$J;d$,label.dist=-0.3mm}{o1}
\fmfv{label=$j;b$,label.dist=-0.3mm}{o2}
\fmf{dashes}{i1,v1}
\fmf{gluon}{i2,v1}
\fmf{dashes}{o1,v1}
\fmf{gluon}{o2,v1}
\end{fmfgraph*}
\end{fmffile}}&\quad=-g_{E}^2\delta_{IJ}(f^{ace}f^{bde}+f^{bce}f^{ade}).
\\\nonumber
\end{align}

\section{Scalar contributions to the one-loop diagrams in Fig.~\ref{fig_oneloop}}
\label{app_detail}
The loop contribution from the SYM scalars to $\delta\cc(\qp)$ comes from the three diagrams
identified in Sec.~\ref{sec_collisionkernel}. We now present the details of their explicit evaluation.
The expressions below give the contribution to the self-energies $\Pi^{00}$ and $\Pi^{zz}$.
\begin{alignat}{2}
\raisebox{-15mm}{
\begin{fmffile}{scalar_1}
\fmfset{curly_len}{2.0mm}
\begin{fmfgraph*}(78,110)\fmfpen{0.5thin}
\fmfstraight
\fmfleft{i1,i2}
\fmfright{o1,o2}
\fmf{vanilla,label.side=left,label=$\qquad A^b_0$,label.dist=2mm}{i1,v1}
\fmf{vanilla}{v1,o1}
\fmf{vanilla,label=$\qquad A^a_0$,label.dist=2mm}{i2,v2}
\fmf{vanilla}{v2,o2}
\fmfv{label=$\bm{p}$,label.dist=15mm}{v3}
\fmffreeze
\fmf{vanilla,label=$\phi^b_J$,label.dist=2.9mm}{v1,v3}
\fmf{vanilla,label=$\phi^a_I$,label.dist=2.9mm}{v3,v2}
\fmfi{dashes}{fullcircle scaled .52w shifted (.76w,.5h)}
\end{fmfgraph*}
\end{fmffile}}\quad
&&
\begin{split}
&=\frac{1}{2}\Big[-g_E^2\delta_{IJ}(f^{ace}f^{bde}+f^{bce}f^{ade}) \Big]\Big[ \int_p \frac{\delta^{cd}\delta_{IJ}}{p^2+m_S^2} \Big]\\
&=-6g_E^2C_A\delta^{ab}\int_p\frac{1}{p^2+m_S^2},
\end{split}
\\
\raisebox{-15mm}{\begin{fmffile}{scalar_2}
\fmfset{curly_len}{2.8mm}
\begin{fmfgraph*}(78,110)\fmfpen{0.5thin}
\fmfstraight
\fmfleft{i1,i2}
\fmfright{o1,o2}
\fmf{vanilla,label.side=left,label=$\qquad A^b_j$,label.dist=2mm}{i1,v1}
\fmf{vanilla}{v1,o1}
\fmf{vanilla,label=$\qquad A^a_i$,label.dist=2mm}{i2,v2}
\fmf{vanilla}{v2,o2}
\fmfv{label=$\bm{p}$,label.dist=15mm}{v3}
\fmffreeze
\fmf{gluon,label=$\phi^b_J$,label.dist=3mm}{v1,v3}
\fmf{gluon,label=$\phi^a_I$,label.dist=3mm}{v3,v2}
\fmfi{dashes}{fullcircle scaled .52w shifted (.76w,.5h)}
\end{fmfgraph*}
\end{fmffile}}\quad
&&
\begin{split}
&=\frac{1}{2}\Big[-g_E^2\delta_{zz}\delta_{IJ}(f^{ace}f^{bde}+f^{bce}f^{ade}) \Big]\Big[ \int_p\frac{\delta^{cd}\delta_{IJ}}{p^2+m_S^2} \Big]\\
&=-6g_E^2C_A\delta^{ab}\int_p\frac{1}{p^2+m_S^2},
\end{split}
\\
\raisebox{-17mm}{\begin{fmffile}{scalar_3}
\fmfset{curly_len}{2.8mm}
\begin{fmfgraph*}(78,110)\fmfpen{0.5thin}
\fmfstraight
\fmfleft{i1,i2}
\fmfright{o1,o2}
\fmf{vanilla,label.side=left,label=$\qquad A^b_j$,label.dist=2mm}{i1,v1}
\fmf{vanilla}{v1,o1}
\fmf{vanilla,label=$\qquad A^a_i$,label.dist=2mm}{i2,v2}
\fmf{vanilla}{v2,o2}
\fmffreeze
\fmf{gluon,label=\vspace{2.6em}$\hspace{-4.6em}\qquad \phi^e_K \qquad \phi^f_L$}{v1,v3}
\fmf{dashes,left,tension=0.35,tag=1,label=$\bm{q_\perp-p}$}{v3,v4}
\fmf{dashes,left,tension=0.35,tag=2,label.side=left,label=$\bm{p}$}{v4,v3}
\fmf{gluon,label=\vspace{-2.3em}$\hspace{-4.6em}\qquad \phi^c_I \qquad \phi^d_J$}{v4,v2}
\end{fmfgraph*}
\end{fmffile}}\quad
&&
\begin{split}
&=\frac{1}{2}\int_p\Big[ig_E\delta_{IJ}f^{acd}(2p-q_\perp)_z	 \Big] \Big[\frac{\delta^{df}\delta_{JL}}{p^2+m_S^2} \Big] \\
&\times\Big[ig_E \delta_{KL}f^{bef}(q_\perp-2p)_z) \Big] \Big[\frac{\delta^{ce}\delta_{KI}}{(q_\perp-p)^2+m_S^2} \Big]\\
&=6g_E^2C_A\delta^{ab}\int_p\frac{2p_z^2}{(p^2+m_S^2)((q_\perp-p)^2+m_S^2)},
\end{split}
\end{alignat}
where in dimensional regularization $\int_p\equiv\int d^dp/(2\pi)^d$, with $d\to3$, and
\begin{align}
&\int_p\frac{1}{p^2+m^2}=-\frac{m}{4\pi},\\
&\int_p\frac{p_z^2}{(p^2+m^2)((p-q_\perp)^2+m^2)}=-\frac{(4m^2+q_\perp^2)\tan^{-1}(q_\perp/2m)+2mq_\perp}{32\pi q_\perp}.
\end{align}
Upon inserting the expressions above in the propagators $G_{00}^2$ and $G_{zz}^2$, 
the contribution of the SYM scalar fields to NLO collision kernel sums up to Eq.~\eqref{scalarcontrib}.

\bibliographystyle{JHEP}
\bibliography{nlocollinearphotonbib}
\end{document}